\documentclass[floatfix, noshowpacs, preprintnumbers, onecolumn, amsmath, amssymb, aps, superscriptaddress, prb]{revtex4}

\pdfoutput=1

\usepackage{graphicx, xcolor}
\usepackage[caption=false]{subfig}
\usepackage{soul}
\usepackage{titlesec}

\usepackage{mathdots}
\usepackage{longtable}
\usepackage{setspace}

\newcommand{\ket}[1]{\left| #1 \right>}

\begin{document}

\title{FeynmanPAQS: A Graphical Interface Program for\\ Photonic Analog Quantum Computing}

\author{Hao Tang}
\email{htang2015@sjtu.edu.cn}
\affiliation{State Key Laboratory of Advanced Optical Communication Systems and Networks, School of Physics and Astronomy, Shanghai Jiao Tong University, Shanghai 200240, China}
\affiliation{Institute of Natural Sciences, Shanghai Jiao Tong University, Shanghai 200240, China}
\affiliation{Synergetic Innovation Center of Quantum Information and Quantum Physics, University of Science and Technology of China, Hefei, Anhui 230026, China}

\author{Yan-Yan Zhu}
\affiliation{School of Physical Science, University of Chinese Academy of Science, Beijing 100049, China}

\author{Jun Gao}
\affiliation{State Key Laboratory of Advanced Optical Communication Systems and Networks, School of Physics and Astronomy, Shanghai Jiao Tong University, Shanghai 200240, China}
\affiliation{Synergetic Innovation Center of Quantum Information and Quantum Physics, University of Science and Technology of China, Hefei, Anhui 230026, China}

\author{Marcus Lee}
\affiliation{Deparment of Physics, Cambridge University, Cambridge CB3 0HE, UK}

\author{Peng-Cheng Lai}
\affiliation{State Key Laboratory of Advanced Optical Communication Systems and Networks, School of Physics and Astronomy, Shanghai Jiao Tong University, Shanghai 200240, China}

\author{Xian-Min Jin}
\email{xianmin.jin@sjtu.edu.cn}
\affiliation{State Key Laboratory of Advanced Optical Communication Systems and Networks, School of Physics and Astronomy, Shanghai Jiao Tong University, Shanghai 200240, China}
\affiliation{Institute of Natural Sciences, Shanghai Jiao Tong University, Shanghai 200240, China}
\affiliation{Synergetic Innovation Center of Quantum Information and Quantum Physics, University of Science and Technology of China, Hefei, Anhui 230026, China}

\email{xianmin.jin@sjtu.edu.cn} 

\maketitle
\textbf{We present a user-friendly software for photonic analog quantum computing with an installable MATLAB package and the graphical user interface (GUI) that allows for convenient operation without requiring programming skills. Arbitrary Hamiltonians can be set by either importing the waveguide position files or manually plotting the configuration on the interactive board of the GUI. Our software provides a powerful approach to theoretical studies of two-dimensional quantum walks, quantum stochastic walks, multi-particle quantum walks and boson sampling, which may all be feasibly implemented in the physical experimental system on photonic chips, and would inspire a rich diversity of applications for photonic quantum computing and quantum simulation. We have also improved algorithms to ensure the calculation efficiency of the software, and provided various panels and exporting formats to facilitate users for comprehensive analysis of their research issues with abundant theoretical results.}
\section*{\normalsize I. Introduction}

Applying quantum simulation to real physical and computational problems has been a main thought of quantum information science since Feynman raised the concept of quantum computing\cite{Feynman1982}. Quantum simulation is to use the Hamiltonian of a quantum system to simulate the Hamiltonian of the target system. The mapping needs not to be highly strict, but only to be able to produce some key features of the target system, and even some qualitative results instead of full quantitative details are very valuable\cite{Buluta2009, Georgescu2014}. In the two major genres of quantum computing, the universal (or digital) one and the analog one, the former is more prone to the influence of errors and rely more heavily on error corrections. On the other hand, the analog quantum computing have the advantages of the lower resource requirements and the higher tolerance level to imperfections of the quantum system. Together with the aforementioned wide demand and loose requirement for simulating target systems, all these facts have made analog quantum computing an important tool for quantum simulation, with a rich diversity of applications in condensed-matter physics, high-energy physics, atomic physics, quantum chemistry, and biology, etc\cite{Buluta2009, Georgescu2014, Arguelloluengo2018, Lambert2013}.

Among a variety of physical systems that have been employed to quantum computing, including solid state devices, atomic or nuclear spin systems, superconducting devices, etc, photons have many inherent advantages for quantum information, due to their fast speed and a lack of the interaction with the environment\cite{Flamini2018, OBrien2010}. In particular, photonic systems are very suitable for analog quantum computing and quantum simulation \cite{Aspuru2012} because the high mobility of photons enables flexible constructions of the corresponding Hamiltonian systems using the photonic systems. Some most representative examples of photonic analog quantum computing include quantum walks and boson sampling that evolves continuously in the well-designed photonic arrays. A quantum walk based on real two-dimensional space was recently demonstrated on a photonic chip\cite{Tang2018}. Quantum walks are regarded as a highly versatile approach for quantum simulation of different tasks, and recent advances for flexible evolution paths on the photonic chip make their real applications more possible\cite{Tang2018, Tang2018b}.  Boson sampling, a scheme that requires only passive linear optics interferometer, single photon source and photodetection, may set a key milestone in the quantum computing field called quantum supremacy\cite{Harrow2017} and have inspired many early experimental explorations\cite{Spring2013, Broome2013, Tillmann2013, Spagnolo2014, Wang2017}.

Along with the advances of quantum information processing devices, quantum software is being rapidly developed, at a phase similar to the early period of computer history when hardware and software assisted each other to move on from the initial stirring of electronic computing machines \cite{Ying2018}. Recent years saw the springing up of a growing number of quantum software with different functions \cite{Finke2018, Quantiki2018, LaRose2018}. A layered software architecture for quantum computing design tools was proposed in 2006 that defined the four-phase design flow to connect the front end to the physical quantum devices, through Quantum Intermediate Representation, Quantum Assembly Language and Quantum Physical Operations Language \cite{Svore2006}. This layered structure inspired a series of software in different classical program languages for universal quantum computing in the past few years, including L$|$QUi$|$$>$\cite{Wecker2014} and Q\# by Microsoft\cite{Microsoft2017,Svore2018}, QISkit by IBM\cite{IBM2018}, and Forest by Regetti\cite{Rigetti2018, Polloreno2018}, a start-up company, as well as ProjectQ by ETH\cite{Steiger2016} and $Q|SI\rangle$ by University of Technology Sydney\cite{Liu2017}. They suppose their software to work once the real universal quantum computers are very powerful and robust. Meanwhile, some quantum clouds were recently launched by IBM \cite{IBMcloud2016} and Alibaba\cite{Alibaba2017} that each have a connection to a real superconducting quantum computer, and by Tsinghua University based on a nuclear magnetic resonance (NMR) computer \cite{Xin2018}. These cloud and software make the pioneering attempt with a focus on universal quantum computing.

Software for analog quantum computing and simulation is the other major part of quantum software. OpenFermion is an open-source analog quantum simulation software developed by Google \cite{Mcclean2017} and specializes in simulating fermionic systems and quantum chemistry. As open quantum systems are always involved in quantum simulation, the software package solving Lindblad master equation for quantum stochastic walks in open quantum systems have emerged, including QuTiP, a Python package \cite{Johansson2012, Johansson2013}, QSWalk, a Mathematica package useful for Hamiltonian based on graphs\cite{Falloon2017}, and some packages tailored to specific tasks of quantum stochastic walks such as the centrality test \cite{Izaac2016, Izaac2017}. These analog quantum packages do not directly correspond to a real quantum physical system to instruct experimental implementation. An exception is Strawberry Fields, a specialized platform for photonic quantum computing, but it focuses on continuous-variable quantum computing only\cite{Killoran2018}. For the more general applications of photonic analog quantum computing, e.g., single and multi-photon quantum walks and boson sampling, there's a `Bristol Quantum Cloud for Boson Sampling', which yet allows for the calculation of a very small scale\cite{Bristol2018}. Overall, there is a lack of software that provides practical instruction for photonic experiments in analog quantum computing.

Therefore, we launch the FeynmanPAQS, a first software for photonic analog quantum computing and simulation. The name is to salute to Feynman for his pioneering ideas on quantum computers and insightful emphasis on quantum simulation\cite{Feynman1982}. PAQS is an acronym for Photonic Analog Quantum Simulation that suggest two key points of the software: the focus on analog quantum computing and the corresponding physical platform on the photonic system. FeynmanPAQS can be compatible with various integrated photonic quantum circuits regardless of their fabrication method, such as Silicon-on-Insulator\cite{Wang2014}, Silica-on-Silicon\cite{Politi2008}, UV writing\cite{Spring2013} and femtosecond laser writing\cite{Feng2016, Szameit2007, Crespi2013, Chaboyer2015}, etc.  The advantages of this software can be summarized as follows:

\begin{itemize}
\item Easy to install the software from the .exe file, and highly convenient to operate using the graphic user interface (GUI), which is generally considered as a very user-friendly approach\cite{Giorgino2017,Umansky2012,Vergaraperez2016}.
\item Supports arbitrary design of the photonic array and the corresponding Hamiltonian. Especially, the interactive board is enabled so that the user could locate the waveguide in any position just by clicking the mouse, besides an alternative way of importing the position information from an excel file.
\item Provides a unique way of realizing open quantum systems that's especially suitable for the photonic system. It inspires analog quantum simulation using the photonic system for various applications.
\item Uses optimization algorithms to speed up the calculation time for multi-particle quantum walks and boson sampling. Multiple panels are also provided to view data from different perspectives.
\end{itemize}

\begin{figure}[h!]
\includegraphics[width=0.5\textwidth]{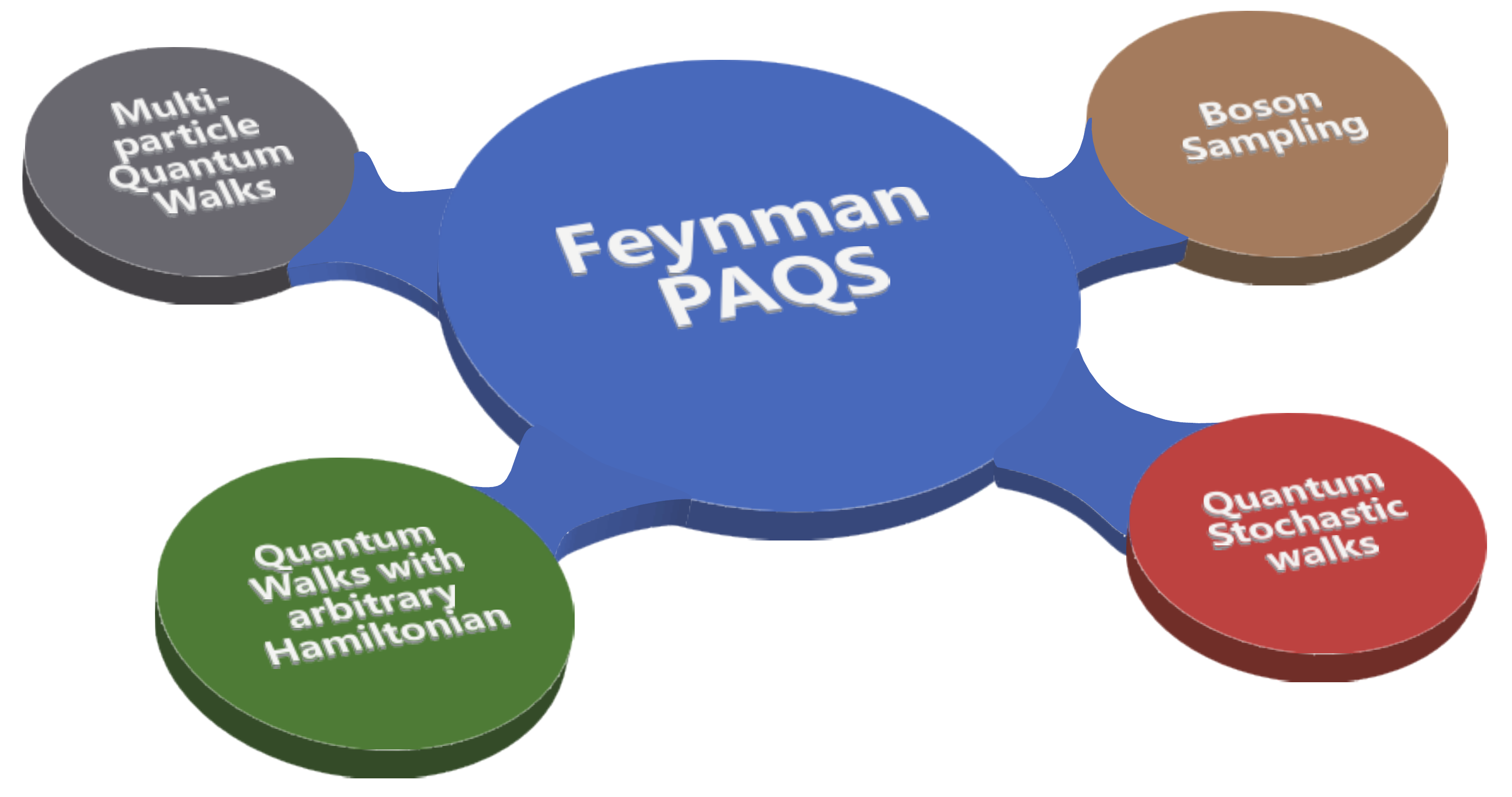}
\caption{\textbf{Framework of FeynmanPAQS.} The software contains four modules, namely, the two-dimensional quantum walks with abitrary Hamiltonian design (QW), quantum stochastic walks on the photonic chip (QSW), multi-particle quantum walks (MultiParticle), and boson sampling (BosonSampling). }
\label{fig:strutturaChip}
\end{figure}

The structure of this paper is as follows. In Section II-V we introduce the four modules of this software, respectively. They are, namely, module for the two-dimensional quantum walks with abitrary Hamiltonian design (QW), quantum stochastic walks on the photonic chip (QSW), multi-particle quantum walks (MultiParticle), and boson sampling (BosonSampling), as shown in Fig.~1. In each section we begin with a brief overview of the theoretical models that support each module. We then introduce some special features of each module that distinguishes them from other present software that has similar functions.  For the module of MultiParticle and BosonSampling, we compare our performance using the `Ryser+Gray \& Glynn+Gray' mixed algorithm with those using other or no optimization algorithms. We further give some case studies to show how to utilize the user interface to conduct the research of quantum physics problems. Then Secton VI gives a brief discussion of this software for photonic analog quantum computing. A full list of all the functions and buttons shown in each module of this software is provided in Appendix.

\section*{\normalsize II. Module for two-dimensional quantum walks with abitrary Hamiltonian design (QW)}

Quantum walks, the quantum analogue of classical random walks \cite{Aharonov1993, Childs2002}, have their unique features of interference and superposition, which lead to remarkly different behaviours and normally faster performances comparing to classical random walks. Therefore, quantum walks become a highly powerful approach to quantum algorithms\cite{Ambainis2003, Shenvi2003, Childs2004}, and quantum simulation for various systems\cite{Mulken2011, Lambert2013, Aspuru2012}. In order to really apply quantum walks to a wide range of applications, a two-dimensional evolution space of a large scale and flexible design of the Hamiltonian for quantum walks are two preliminary requirements. A few examples of such large-scaled two-dimensional quantum walks have now been experimentally realized on photonic chips\cite{Tang2018, Tang2018b}. This module of QW provides a platform to allow for more arbitrary Hamiltonian designs to help researchers explore a rich diversity of quantum walk simulation applications.

As concerned in this module (QW), for two-dimensional quantum walks with single photons injected into the photonic chip consisting of straight waveguides, photons propagating through these coupled waveguides can be described by the Hamiltonian:
$$H=\sum_{i}^N \beta_i a_i^\dagger a_i + \sum_{i \neq j}^N C_{i,j} a_i^\dagger a_j\eqno{(1)}$$
where $\beta_i$ is propagating constant in waveguide $i$, $C_{i,j}$ is the coupling strength between waveguide $i$ and $j$. $C_{i,j}$ that mainly depends on waveguide spacing can be experimentally obtained via a coupled mode approach \cite{Szameit2007}. In the module, we use the empirically fitted relationship between the coupling coefficient $C$ (unit: $\rm cm^{-1}$) and the waveguide spacing $d$ (unit: $\mu$m) as follows:  $C=41.42\times \exp(-d/4.616)$. Hence once the waveguide spacing is set, the coupling coefficient in the Hamiltonian is obtained (see Fig.~2).

\begin{figure}[h!]
\includegraphics[width=0.7\textwidth]{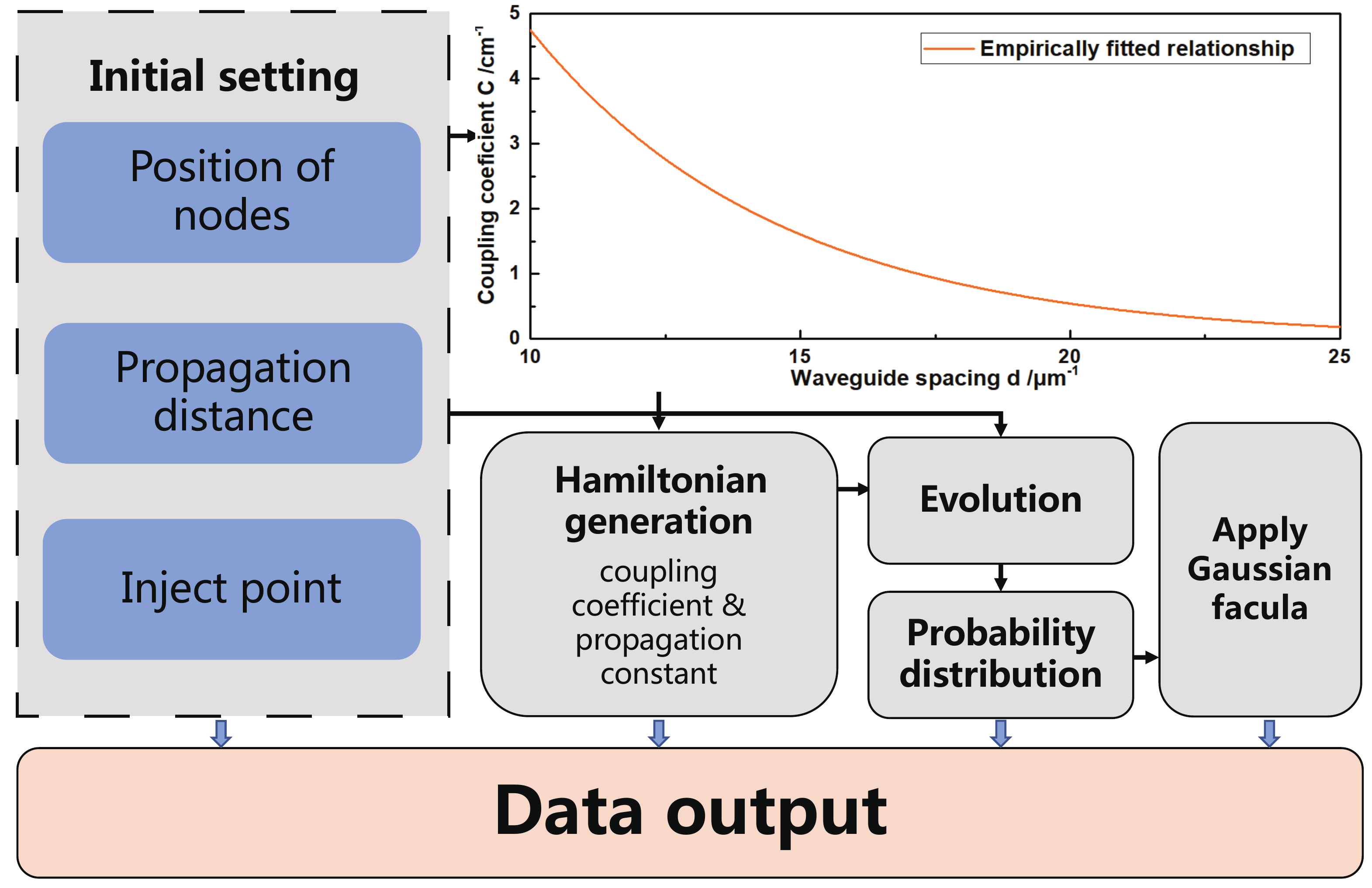}
\caption{\textbf{Theoretical calculation for quantum walks on a photonic chip.} The coupling coefficient $C$ as a function of the centre-to-centre waveguide spacing, and the flow chart for the procedures of calculating the evolution result for quantum walks on a photonic chip.
}
\label{fig:QWflowchart}
\end{figure}

For a quantum walk that evolves along the waveguide, the propagation length $z$ is proportional to the propagation time by $z = ct$, where $c$ is the speed of light in the waveguide, so we use $z$ instead of  $t$ for simplicity. The wavefunction that evolves from an initial wavefunction satisfies:
$$\ket{\Psi(z)}=e^{-iHz}\ket{\Psi(0)}\eqno{(2)}$$
where $\ket{\Psi(z)}=\sum_ja_j(z)\ket{j}$, and $|a_j(z)|^2=P_j(z)$. $|a_j(z)|^2$ and $P_j(z)$ is the probability of the walker being found at waveguide $j$ at the propagation length $z$.

In experiment, we would observe such probability distribution by injecting a laser beam or single photon source into one waveguide and measuring the evolution patterns using a certain type of CCD camera. The normalized light intensity of each waveguide corresponds to the probability at this waveguide and the facula at each waveguide is normally in a shape of the two-dimensional Gaussian distribution. In the software module, in order to give better theoretical instruction for photonic quantum walk experiment, we plot the same Gaussian-shaped facula for presenting the probability distribution.

In short, the workflow in this module for theoretically calculating quantum walks on a photonic chip mainly includes the following four steps: Generate Hamiltonian, Evolve(Exponentiate Hamiltonian), Obtain probability distribution and Apply Gaussian facula (See Fig.~2). A GUI of this module is further given in Fig.~3 showing a case of a two-dimensional quantum walk with an arbitrary Hamiltonian design from the panel `\textsf{Input Your Own One}'. The workflow of the calculation process can be revealed in these panels on GUI. Detailed explanation of the functions for every button in each panel is provided in Appendix.

\begin{figure}[t!]
\includegraphics[width=1.0\textwidth]{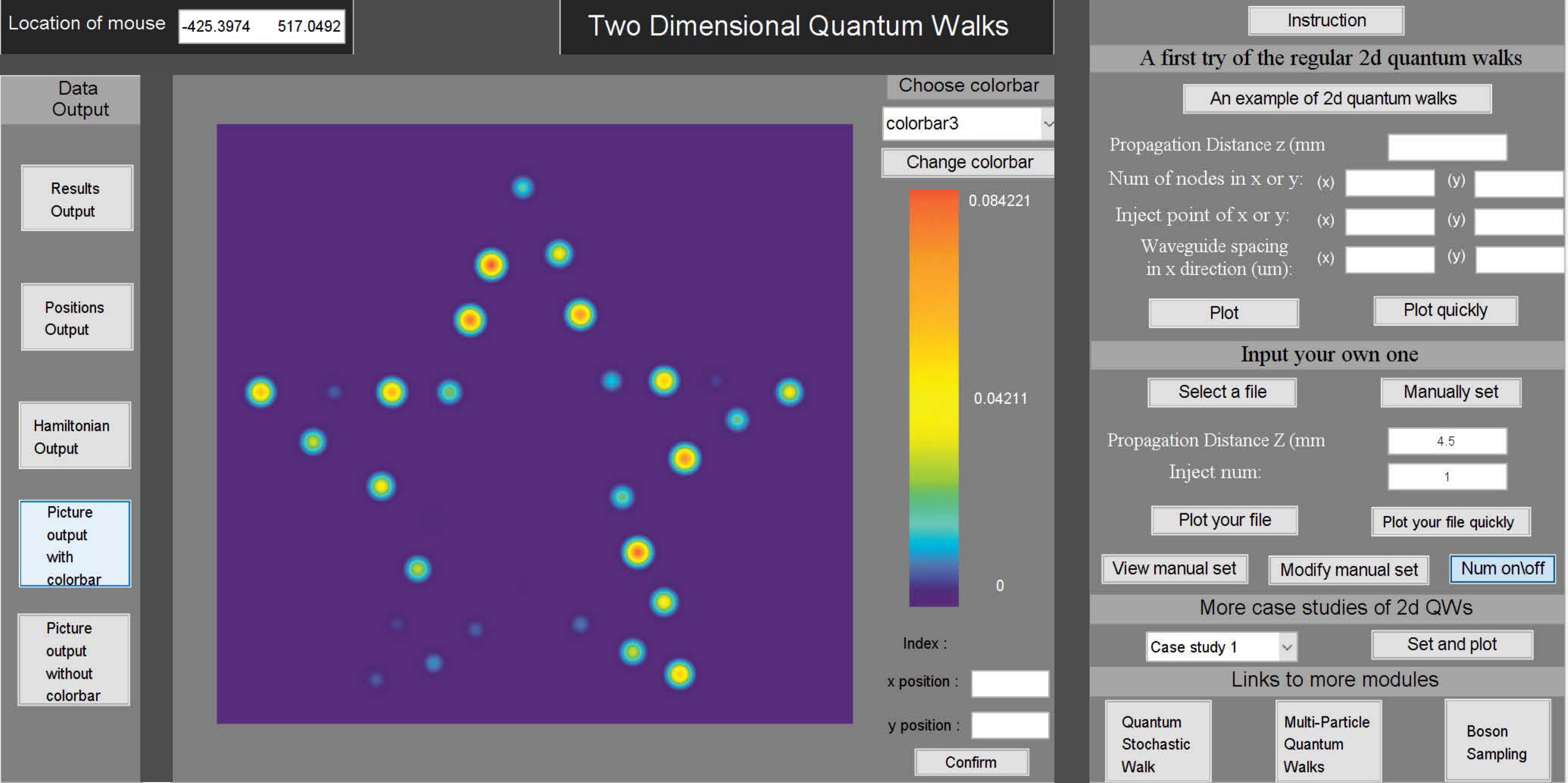}
\caption{\textbf{The MATLAB graphical user interface for the module of QW.} An abitrary array forming the shape of a pentagram has been set by manually plotting the waveguide in the interactive board, and its evolution pattern with Gaussian-shaped facula for a certain propagation distance and photon-injected waveguide No. is plotted. 
}
\label{fig:GUIforQW}
\end{figure}

\section*{\normalsize III. Module for quantum stochastic walks on a photonic chip (QSW)}

Real systems often don't evolve fully according to the pure quantum model, as environmental noise easily causes some degree of decoherence, leading to the very common open quantum system issues. Lots of quantum simulation researches aim at addressing the issues of open quantum systems\cite{Georgescu2014, Arguelloluengo2018, Lambert2013}, and the quantum stochastic walk \cite{Whitfield2010} is considered a very useful tool.

The most common formulation of the quantum stochastic walks uses the Lindblad master equation, which is a differential equation that incorporates both quantum and classical walk in the continuous-time evolution\cite{Whitfield2010, Falloon2017}:

$$\frac{d\rho}{dt}=-(1-\omega)i[H,\rho]+\omega\sum_{ij}(L_{ij} \rho L_{ij}^\dagger- \frac{1}{2}\{L_{ij}^\dagger L_{ij}, \rho\}).\eqno{(3)}$$
The first part right to Eq.~(3) represents the quantum walk evolution, where $H$ is the Hamiltonian operator. The second part, which contains the Lindblad operators $L_{ij}$, describes the classical random walk evolution. The parameter $\omega$ interpolates the weight of quantum walk and classical walk, that is, a higher value of $\omega$ suggests a larger portion of the classical walk in the mixture of quantum stochastic walk.

There are a few software that can be used to solve Lindblad equations, such as Qutip\cite{Johansson2012, Johansson2013} and QSWalk\cite{Falloon2017}.  However, the parameters in the Lindblad equation does not have a clear correspondence to the physical parameters of the photonic quantum systems. In the photonic chip, it is difficult to directly realize the Lindblad equation, and instead, a $\Delta \beta$ photonic model was raised\cite{Caruso2016} to achieve quantum stochastic walks in the photonic chip with continuous-time evolution paths:

$$ i\frac{d\psi_n}{dz} = \Delta \beta_n(z)\psi_n+\sum_{m\neq n}C_{mn}\psi_m \eqno{(4)}$$
where $\psi_n$ is the wave function at waveguide $n$, and $\Delta \beta_n(z)$ is the change to the propagation constant $\beta_n(z)$. The $\beta_n(z)$s are originally a constant if the photonic chip fabrication parameters are kept stable, and the photonic array would be a purely quantum evolution system for quantum walks. However, when $\Delta \beta_n(z)$s as the fluctuation of the propagation constant are added along the evolution path, which can be physically realized by randomly tuning certain fabrication parameters along the waveguide (see Fig.~4), the noise is introduced into the photonic quantum system. The strength of the noise is positively related to the amplitude of the random fluctuations by $\Delta \beta_n(z)$s.

The experimental $\Delta \beta$ approach is not a strict mapping of the Lindblad master equations, but they both generate very similar effects. The mechanism of introducing the environment decoherence by the Lindblad terms lies in that, the Lindblad terms added fluctuations in the diagonal directions of the probability matrix. In the $\Delta \beta$ model, by adding random values of $\Delta \beta$ along the waveguides, fluctuations are also generated in the diagonal position of the Hamiltonian matrix to serve as the environment decoherence.

The GUI for this module of QSW is very similar to that for the module of QW, in that they both allow for the abitrary two-dimensional array design and follow the same four-step workflow to get the calculation result. Still, QSW has its special buttons and panels to facilitate the study of $\Delta \beta$ approach for quantum stochastic walks. Users can define how frequently  $\Delta \beta$ varies along the waveguide (by filling `\textsf{z interval}'), which ones of the waveguides need to add $\Delta \beta$ variation (by choosing `\textsf{Stochastic or not}' ) and how many times of random settings to get an average result (by filling `\textsf{Average of X times}'), etc. An extra $\rho_z$ panel showing the evolution probability against the evolution length is also provided. Quantum stochastic walks tend to have less dynamic evolution than pure quantum walks, and the stationary probability at certain waveguides may convey some physical meaning, such as energy transport efficiency, so this $\rho_z$ panel as well as the corresponding exporting data file may be helpful.

\begin{figure}[t!]
\includegraphics[width=0.5\textwidth]{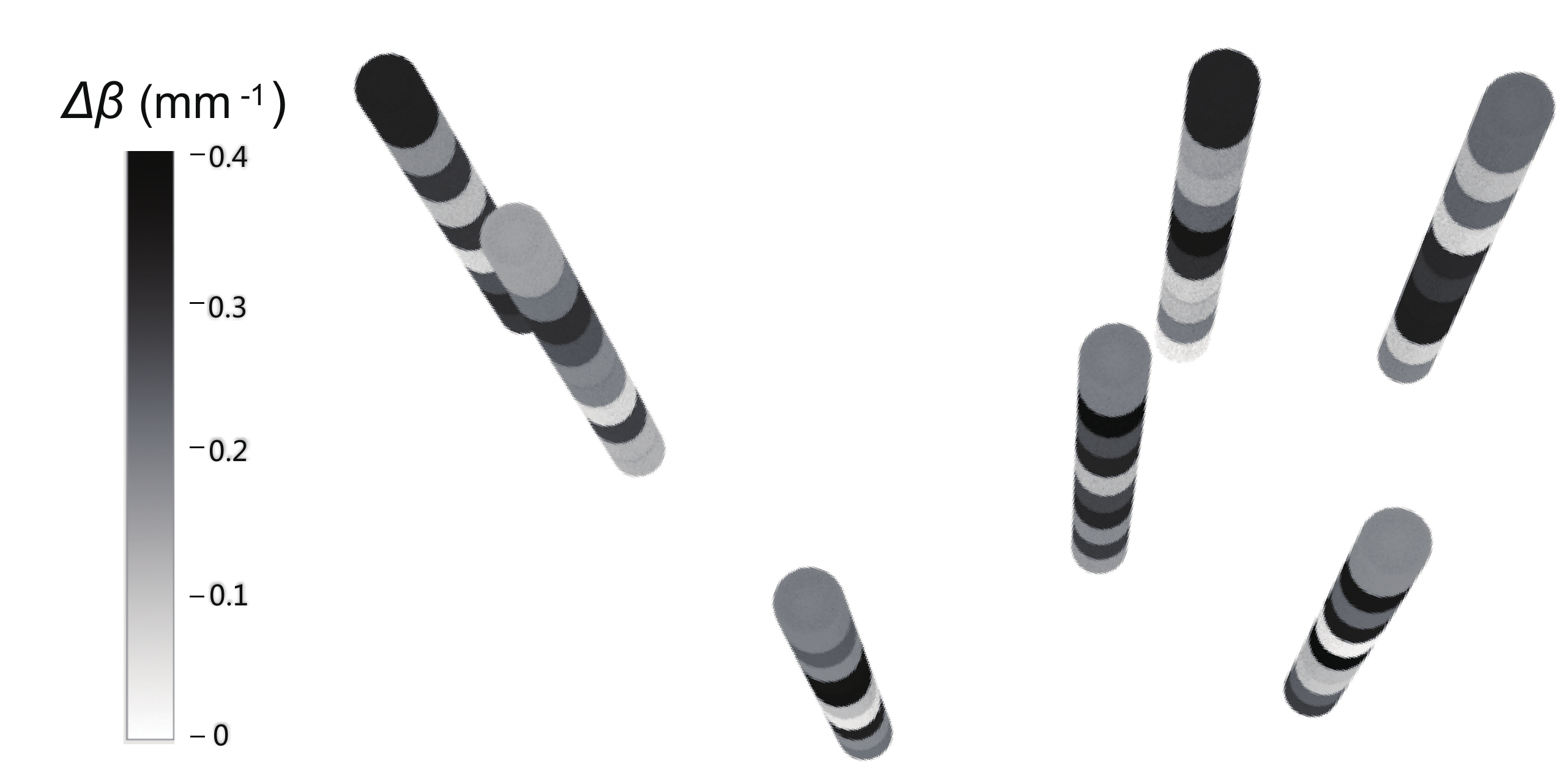}
\caption{\textbf{The $\Delta \beta$ photonic model.}  Schematic diagram of the randomly varying propagation constants shown in different colors along the propagation direction of an abitrary array formed by seven waveguides. This set of random values is one example of the case with a $\Delta \beta$ amplitude of 0.4~$\rm mm^{-1}$.}
\label{fig:QSW}
\end{figure}

\section*{\normalsize IV. Module for multi-particle quantum walks (MultiParticle)}
So far we have been focusing on a single particle evolving in a two dimensional quantum system. It is natural to wonder how the dynamics can be richer if we inject more than one particle into the system, i.e two or more indistinguishable particles. In this case, two main features of quantum optics, quantum interference and quantum correlation play central roles. The genuine quantum interference (also known as Hong-Ou-Mandel (HOM) effect\cite{Hong1987}), which cannot be classically simulated, gives rise to non-classical quantum correlations\cite{Peruzzo2010}. With two particles populating in a two dimensional lattice, the corresponding state space can be easily extended to a greater dimension\cite{Gao2016}. Such high dimensional graphs demonstrate quantum speedup for continuous-time search algorithm\cite{Childs2004}.

Another striking departure of quantum mechanics from classical physics is that the particle statistics, either bosonic or fermionic, can influence the quantum correlation, exhibiting either bunching or antibunching behaviors. These phenomenon can be physically simulated with bi-partite entanglement in the form of

$$ {\left| \psi  \right\rangle}= \frac{1}{\sqrt{2}}({a_i^\dag }{b_j^\dag }+e^{i\phi}{a_j^\dag }{b_i^\dag }) {\left| 00  \right\rangle} \eqno{(5)} $$ 
Phase $\phi$ controls the states' symmetry, thus determines the particle statistics.

With even greater $N$ particles involved in the evolution through a system of $M$ waveguides, the process can be regarded as a generalized HOM effect and a photon scattering model. The outcome also depends on whether the particles are distinguishable or indistinguishable and if they are indistinguishable, whether they are bosons or fermions. This distinguishability depends on whether the particles are identical, e.g., whether they have temporal delay or the same frequency when they enter the waveguide array.

If the particles are indistinguishable bosons, then given a Hamiltonian $H$ and its corresponding unitary evolution $U=e^{-\frac{iHt}{\hbar}}$, we can calculate the probabilities of various states occurring. If we define a configuration $S_i$ to be the number of photons injected in waveguide $i$, and $T_j$ to be the number of photons exiting waveguide $j$, then the probability of an initial state $S=|S_1\cdot\cdot\cdot S_M>$ evolving into a final state $T=|T_1\cdot\cdot\cdot T_M>$ is given by\cite{Tillmann2013}:

  $$  P(T|S)=\frac{|{\rm Per~} (U^{(S,T)} )|^2}{\sum_{i=1}^M S_i ! \sum_{j=1}^M T_j !} \eqno{(6)} $$
where Per is the permanent of a matrix, and $U^{(S,T)}$ is a submatrix of $U$ that can be constructed by taking $S_i$ copies of the $i^{th}$ column of $U$ and taking $T_j$ copies of the $j^{th}$ row of $U$. Since $\sum_{i=1}^M S_i =\sum_{j=1}^M T_j =N$, $U^{(S,T)}$ has dimensions $N\times N$.
Fermions have a similar formula:

  $$  P(T|S)=\frac{|{\rm Det~}(U^{(S,T)} )|^2}{\prod_{i=1}^M S_i ! \prod_{j=1}^M T_j !} \eqno{(7)}$$
where Det is the determinent of a matrix. As for distinguishable particles, they evolve independently (classically), hence cross terms can be eliminated by taking a linear combination of the determinant and permanent:

  $$  P(T|S)=\frac{1}{2}  \frac{|{\rm Per~}(U^{(S,T)} )|^2+|{\rm Det~}(U^{(S,T) } )|^2}{\prod_{i=1}^M S_i !\prod_{j=1}^M T_j !} \eqno{(8)} $$

With these formulae incorporated into the module of MultiParticle, we are able to use this software to calculate the multi-particle evolution. The GUI for MultiParticle is consistent with that for QW and QSW, as their processes are the same for setting arbitrary waveguide patterns, and the modules all allow for exporting the calculated results in graph and data forms. However, MultiParticle involves many more panels and results. Fig.~5 shows a GUI of this module for a case of injecting three photons (particle type: \textsf{Bosonic}) in a waveguide array, with each single photon injected in Waveguide No. 1, 5 and 9 respectively. The above panel shows the break-down of probability distribution of all states, and a scroll bar can be used for the very long list. The bottom-left panel shows the two-particle coincidence when the other $N$-2 photons have certain locations, which can be set by the pop-up menu `\textsf{View perspectives}'. The bottom-right panel shows the Gaussian-shaped facula for a single photon that can be specified by the pop-up menu `\textsf{View photon No.}'. The panel on the right plotting the probability against the evolution length may be useful for analyzing the bunching and anti-bunching effects for different types of particles\cite{Sansoni2012, Matthews2013}.

\begin{figure}[t!]
\includegraphics[width=1.0\textwidth]{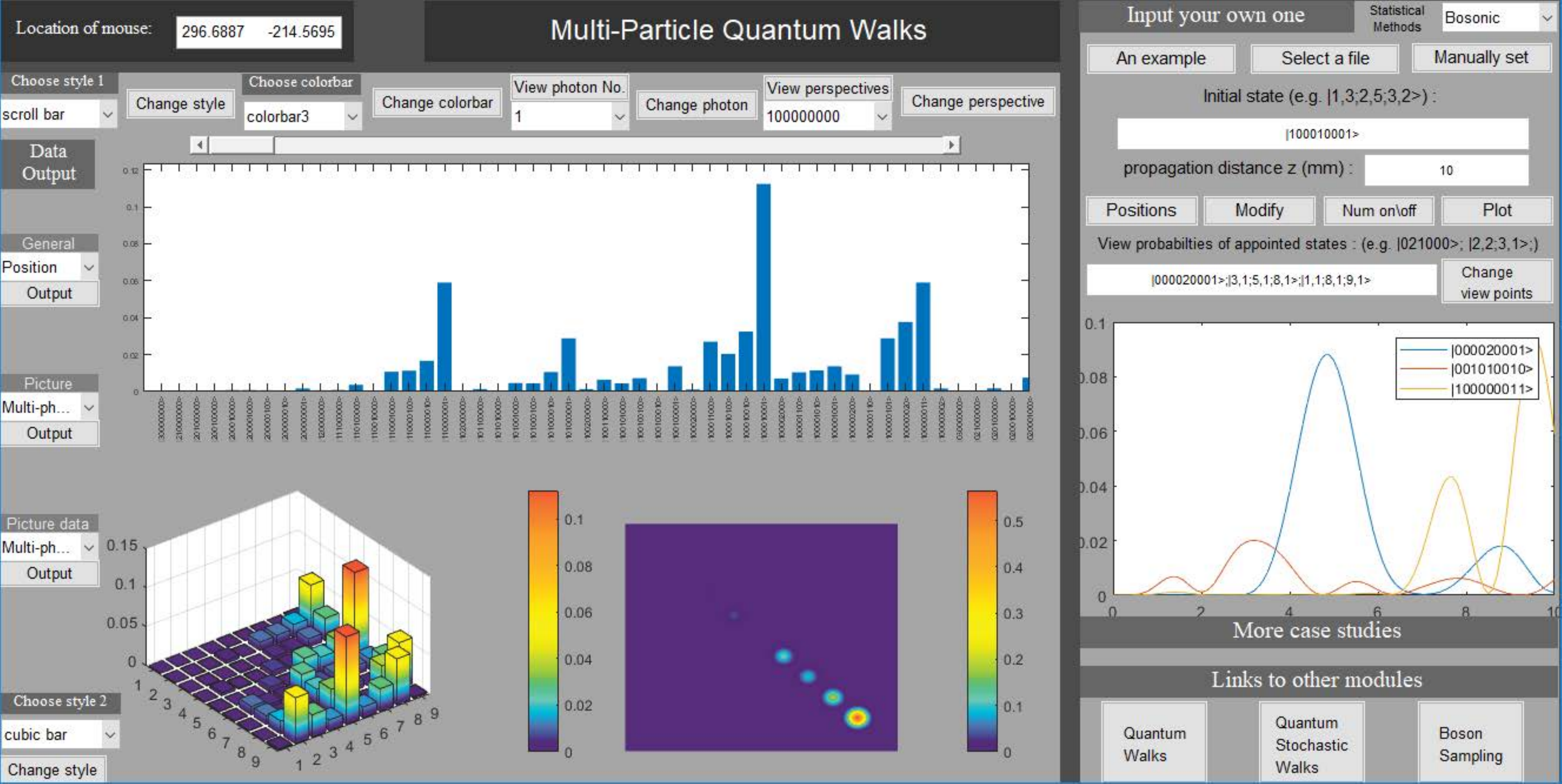}
\caption{\textbf{The MATLAB graphical user interface for the module of MultiParticle.}
}
\label{fig:GUIforMultiParticle}
\end{figure}

It is also worth mentioning that we manage to use effective optimization algorithms for the key part of the theoretical model of this module: the permanent calculation. The formula is of much interest, since computing the permanent has been proven to be of $\#$P complexity, even harder than NP-complete, and is widely believed to be classically intractable. The permanent of an $n$-by-$n$ matrix per $M$ is defined as:

 $$ {\rm Per~} M = \sum_{\sigma \in S} \prod_{i=1}^n M_{i,\sigma (i)} \eqno{(9)} $$
where $\sigma$ is a permutation of the set $\{ 1,\dots,n \}$, and $S$ is the group of all such permutations. If we try to compute the permanent based on this formula alone, we obtain an algorithm with time complexity $O(n!n)$, which is very inefficient. Two better algorithms have been discovered, the first of which is due to Ryser\cite{Ryser1963}:

 $$ {\rm Per~} M = (-1)^n \sum_{\epsilon \in S} (-1)^{\sum_{i} \epsilon_i} \prod_{k=1}^n \sum_{j=1}^n \epsilon_j M_{jk} \eqno{(10)} $$
where $\epsilon = (\epsilon_1, \dots, \epsilon_n) \in S = \{ 0,1 \}^n$ (i.e. $\epsilon$ is a $n$-dimensional vector where all elements are 0 or 1, and $S$ is the set of all $2^n$ possible $\epsilon$). The second algorithm is due to Glynn\cite{Glynn2010}:

$$ {\rm Per~} M = 2^{1-n} \sum_{\delta \in P} \left( \prod_{i=1}^n \delta_i \right) \prod_{k=1}^n \sum_{j=1}^n \delta_j M_{jk}\eqno{(11)} $$
where $\delta = (1, \delta_2, \dots, \epsilon_n) \in P = \{ \pm 1 \}^n$ and there are $2^{n-1}$ possible $\delta$, since we fix $\delta_1 = 1$.

These 2 formulae are both related to polarisation identities of symmetric tensors, and are in fact part of a larger family of permanent algorithms.\cite{Glynn2013}

These formulae give algorithms of complexity $O(2^n n^2)$, but can be reduced to $O(2^n n)$ by the use of a Gray code\cite{Nijenhuis1978}. We will explain this using Ryser's formula, but the same logic applies to Glynn's. If we say that
$\lambda_k = \sum_{j=1}^n \epsilon_j M_{jk}$, then if $\epsilon$ and $\epsilon '$ differ by a single $\epsilon_m$ (i.e. the Hamming distance is 1), then
$\lambda_m ' = \lambda_m + (\epsilon_m ' - \epsilon_m) M_{mk}$, and we only need to calculate one element of the sum instead of all $n$ elements.

Comparing the 2 formulae, it would seem that Glynn's formula is faster by Ryser's by a factor of 2, as the number of possible $\delta$ is half the number of possible $\epsilon$. This is indeed true as $n \rightarrow \infty$, but Ryser's formula seems to be better optimised for small matrices where $N<6$ in our implementations (see Fig.~6).

Therefore, in this software, we use `Ryser+Gray \& Glynn+Gray' mixed algorithm to calculate the permanent, i.e., `Ryser+Gray' for small $N$ ($N<6$) and `Glynn+Gray' for large $N$. This ensures the software to always have an optimized permanent calculation efficiency for scenarios of different photon counts .

\begin{figure}[t!]
\includegraphics[width=0.7\textwidth]{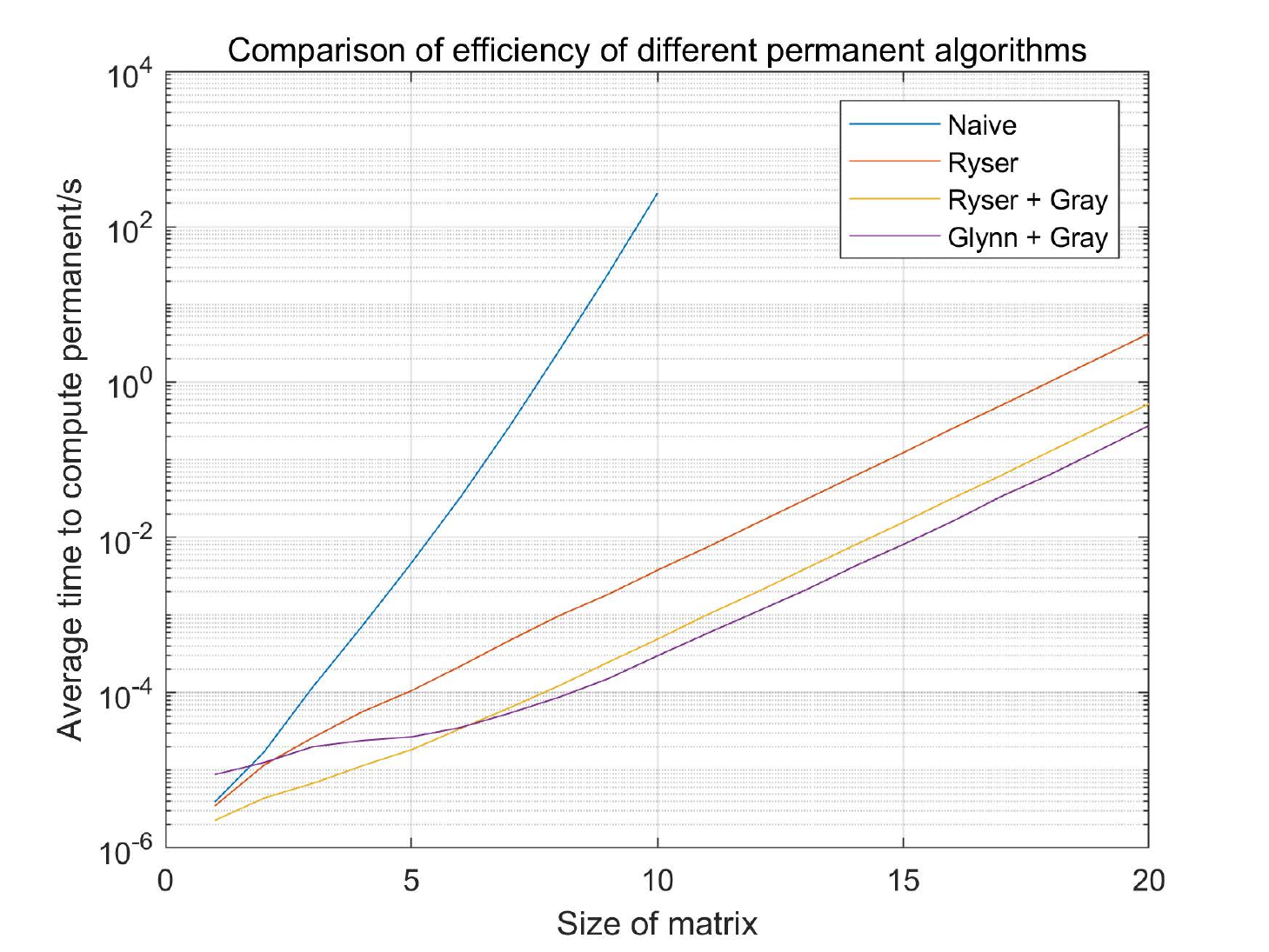}
\caption{\textbf{Comparison of calculation time using different permanent algorithms.}
}
\label{fig:Permanent}
\end{figure}

\section*{\normalsize V. Module for Boson sampling (BosonSampling)}

Boson sampling, a simplified quantum computing model first raised by Aaronson \& Arkhipov\cite{Aaronson2013}, has been becoming an emerging research field in the quantum information community. Unlike universal quantum computation scheme which usually requires active elements, boson sampling scheme only requires passive linear optics interferometer, single photon source and photodetection (photon number resolving is not necessary). Results show that these experimentally friendly devices may offer the first evidence that refute the extended Church-Turing (ECT) thesis and set a key milestone in the quantum computing field called quantum supremacy\cite{Harrow2017}.

The boson sampling problem can be modeled as a multi-photon scattering process. We first prepare an input state consisting of $N$ single photons in an $M$ modes passive linear optics interferometer. The process is very similar with that in the multi-particle quantum walks where $N$ single photons are injected into $M$ waveguides, but a key difference lies in that, there is not a unitary matrix in boson sampling schemes that can be effectively decribed by the Hamiltonian, like $U=e^{-\frac{iHt}{\hbar}}$ in the case of multi-photon quantum walks. In boson sampling, one normally defines $U$, a unitary map on the creation operators, directly.
The injected photons in the boson sampling scheme interfere and scatter in the linear optics interferometers. The interferometer implements a Haar-random $M$-mode transformation $U$ on these $N$ indistinguishable bosons:
$$ U{a_i^\dag }U^\dag=\sum_{j=1}^{M}U_{i,j}{a_j^\dag } \eqno{(12)}$$
where $i$, $j$ denotes the $i$th and $j$th mode of the interferometer. After the transformation, an input configuration which is denoted as $S$ with $\sum_{i=1}^{M}S_i=N$ becomes an output state, which is a superposition of different configurations $T$ for the output modes, denoted as ${\left| \psi_{out}  \right\rangle}=\sum_{T}\gamma_{S,T}{\left| n_1^T n_2^T\cdots\ n_m^T  \right\rangle}$, where $n_i^T$ is the corresponding photon number $n$ in the $i$th mode.

The hardness of this problem roots in evaluating the value of $P(T|S)$, related to the permanent of the scattering amplitudes. The calculation can follow the same equation with Eq.~(6).

\begin{figure}[h!]
\includegraphics[width=1.0\textwidth]{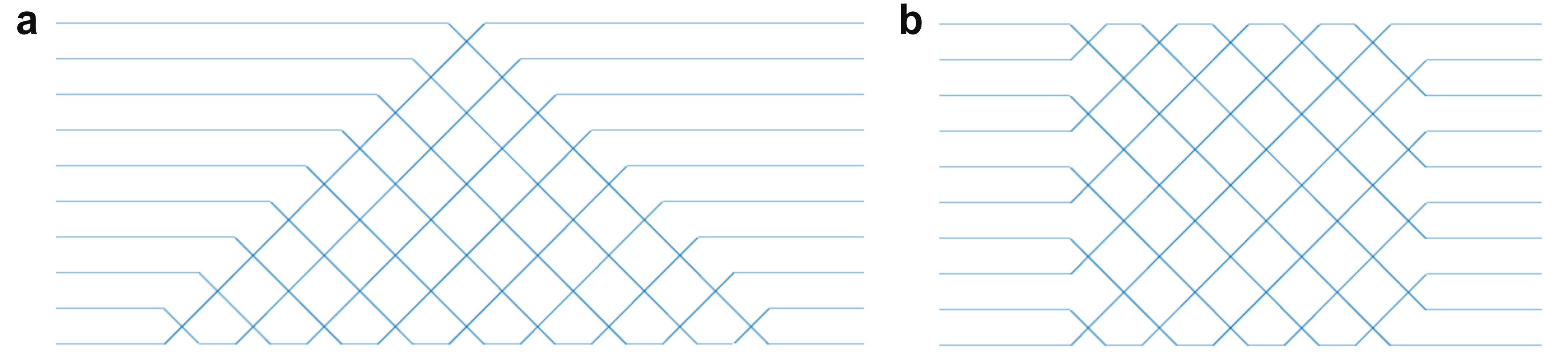}
\caption{\textbf{Schematic diagram of the decomposition.} An example of ({\bf a}) the Reck decomposition with 10 modes, and ({\bf b}) the Clements decomposition of 10 modes.}
\label{fig:BosonSampling}
\end{figure}

\begin{figure}[htbp]
\includegraphics[width=1.0\textwidth]{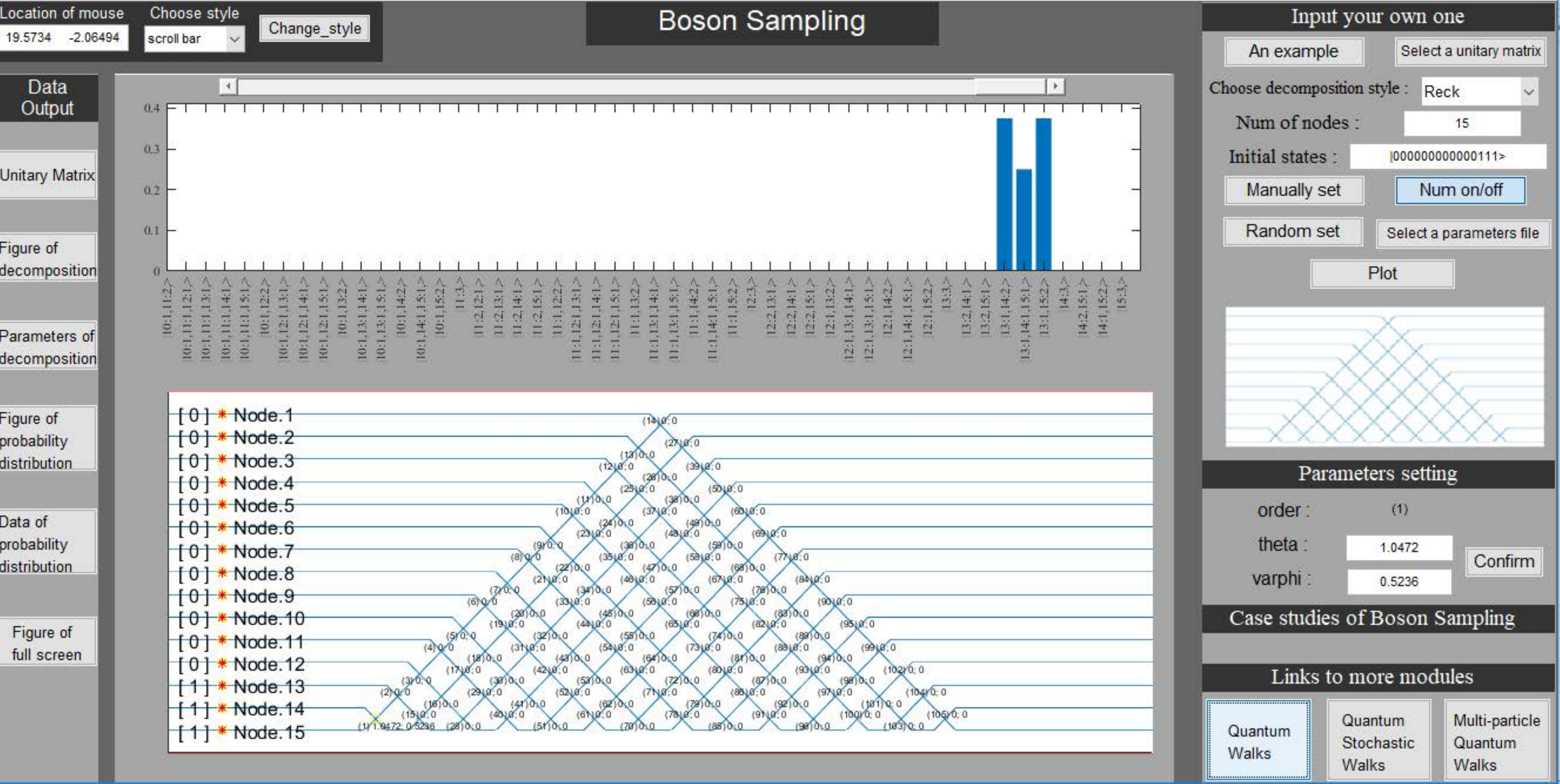}
\caption{\textbf{The MATLAB graphical user interface for the module of BosonSampling.}}
\label{fig:GUIforBosonSampling}
\end{figure}

In the GUI of BosonSampling, users need to import an  $M \times M$ unitary scattering matrix as $U$, and will be informed an error if the imported matrix does not satisfy the requirement for a unitary matrix. The users also need to design their interferometers. Generally, there are two types of decomposition, namely, the Reck's type\cite{Reck1994} and the Clements's type\cite{Clements2016} (see Fig.~7).

After choosing the unitary scattering matrix $U$ and the decomposition style, users would need to set the parameters for beam splitters and phase shifts which accomplish pairwise transformations between channels by satisfying: $U=D (\prod ^N T_{m,n})$, where $D$ is a diagonal matrix and $T_{m,n}$ is transformation from channel $m$ to $n$ ($m=n-1$) through a loseless beam splitter with the reflectivity cos$\theta$, and through a phase shift $\phi$ at input $m$ side\cite{Clements2016}. $T_{m,n}(\theta,\phi)$ reads as:

$$
    T_{m,n}(\theta, \phi) = \left(
    \begin{matrix}
        1       & 0 & \cdots & \cdots & \cdots & \cdots & \cdots & 0 \\
        0       & 1 & \      & \      & \      & \      & \iddots& \vdots \\
        \vdots  & \ & \ddots & \      & \      & \iddots& \     & \vdots \\
        \vdots  & \ & \      & e^{i\phi}\cos{\theta} &-\sin{\theta} & \ & \ &\vdots \\
        \vdots  & \ & \      & e^{i\phi}\sin{\theta} &\cos{\theta} & \ & \ &\vdots \\
        \vdots  & \ & \iddots& \      & \      & \ddots & \      & \vdots \\
        \vdots  & \iddots & \      & \      & \      & \      & 1       & 0 \\
        0       & \cdots & \cdots & \cdots & \cdots & \cdots & 0       & 1 \\
    \end{matrix}
    \right) \eqno{(13)}
$$ \ 

Users could either import the parameters for all $T_{m,n}$s from an excel file or manually type them in the interactive board of the GUI. They could also ask the software to randomly set the parameters of $\theta$ and $\phi$ by choosing `\textsf{Random set}'. A GUI for BosonSampling is shown in Fig.~8, where the decomposition with defined type, number of modes and the injected photons is shown in the interactive board, and the interferometer parameters are marked at the corresponding positions. The obtained probability distribution data can be exported to facilitate further analysis for boson sampling.

\section*{\normalsize VI. Discussion}
Photonic quantum systems can be a very promising physical platform for analog quantum computing and quantum simulation, and so far an increasing amount of efforts are being made for their experimental demonstration. However, there is currently no software that provides comprehensive support for the studies of photonic analog quantum computing and quantum simulation.  We launch FeynmanPAQS in an installable MATLAB package with a very user-friendly GUI, in order to facilitate users from different research fields to work on the multidisciplinary quantum information science and technology based on photonic systems, without requiring them to read complex quantum operators and write programming scripts. We have incorporated the most powerful and versatile tools of analog quantum computing in FeynmanPAQS, covering two-dimensional quantum walks, quantum stochastic walks, multi-particle quantum walks and boson sampling, and these tools can be feasibly implemented in the physical system on photonic chips in experiment. FeynmanPAQS allows for arbitrary Hamiltonian designs and engineering by either importing position files or manually plotting the configuration on the interactive board of the GUI, and allows for flexible setting of input photons, particle types, mode number, interferometer parameters, etc, which provide a highly flexible platform to inspire brainstorming for various simulation proposals and potential applications connecting real problems. We have also improved algorithms to ensure the calculation efficiency in the software, especially for the permanent calculation in the MultiParticle and BosonSampling module, making FeynmanPAQS so far a most efficient software for these tasks to run smoothly on a single laptop. The version of the software described in this paper is FeynmanPAQS 1.0. We will share the software in a cloud server and meanwhile keep updating more case studies of new algorithms and applications to promote the utilization of FeynmanPAQS in the near future.

\bigskip
\textbf{Acknowledgements.}

The authors thank J.-W. Pan for helpful discussions. This research is supported by National Key R\&D Program of China (2017YFA0303700), National Natural Science Foundation of China (11690033, 61734005, 11761141014, 11374211), Science and Technology Commission of Shanghai Municipality (STCSM) (15QA1402200, 16JC1400405, 17JC1400403), and Shanghai Municipal Education Commission (SMEC)(16SG09, 2017-01-07-00-02-E00049). X.-M. Jin acknowledges support from the National Young 1000 Talents Plan.

\newpage
\section*{\normalsize Appendix:}
\subsection{\normalsize Two-dimensional Quantum Walks}
\subsubsection{An overview of the user interface}
\begin{figure}[htbp]
    \centering
    \includegraphics[scale = 0.35]{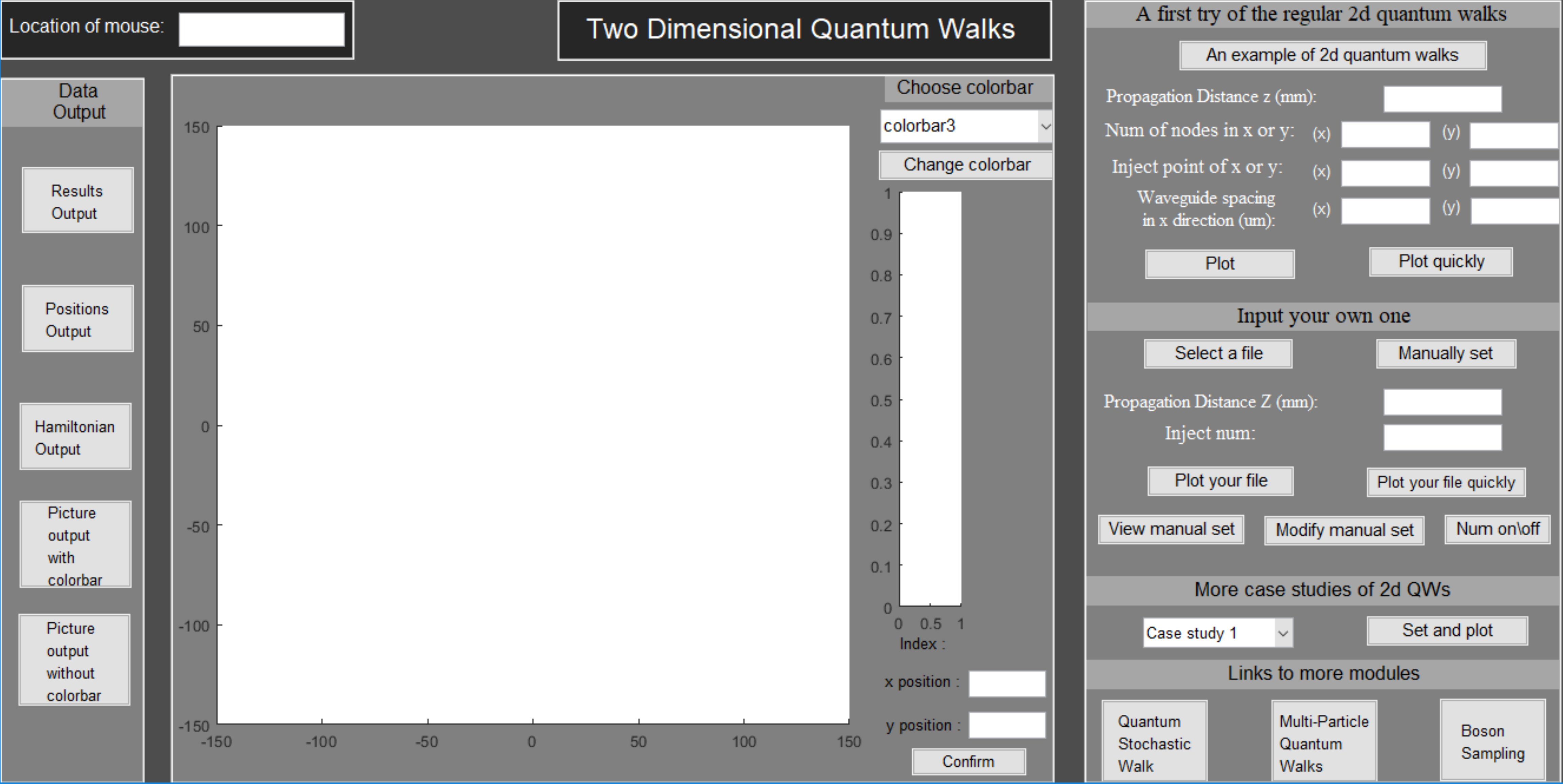}
    \caption{\textbf{The full GUI for the module of QW before any settings.}}
\end{figure}
A full GUI for this module is shown in Fig.~9. There are a few ways to enter waveguide positions. A rectangular array can be selected, and is used in our example setup. Alternatively, there is a manual input function that can select positions based on the position of the cursor, or by entering coordinates. The following will be an element-by-element guide to the program interface. \textsf{}

\subsubsection{Waveguides in a square/rectangular lattice}

\begin{figure}[htbp]
    \centering
    \includegraphics[scale = 0.45]{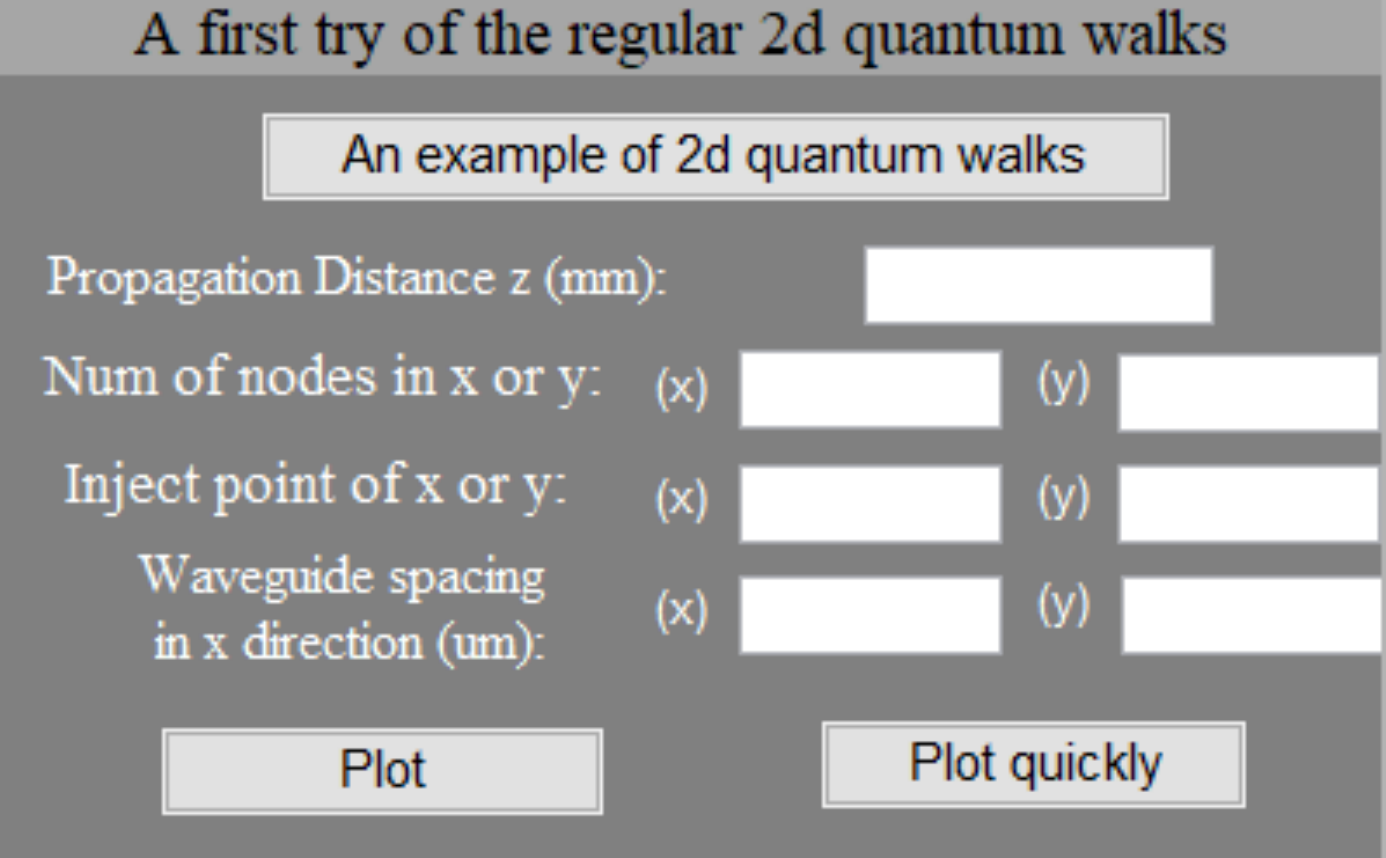}
    \caption{\textbf{The panel for setting a two-dimensional quantum walk on a regular square/rectangular lattice.}}
\end{figure}

\begin{longtable}{p{3cm}p{14cm}}
\ & This subsection explains the elements on the panel as shown in Fig.~10.\\ \\
\textsf{Propagation \qquad\quad distance z}    &    This is the distance propagated along the waveguide array where we simulate measurement. This determines the time allowed for the wavefunction to evolve, through $z = ct$, where $c$ is the speed of light in the waveguide.   \\ \\
\textsf{No. of nodes}    &    The values inputted here control the number of waveguides in the $x$ and $y$ dimensions.   \\ \\
\textsf{Inject point}    &    This controls the point at which the single photon enters the waveguide array. Numbering goes from left to right ($x$-direction), and from top to bottom ($y$-direction). For example, if (1,1) is entered in this field, a photon will be injected at the top left corner.   \\ \\
\textsf{Waveguide spacing}   &    This determines the spacing between the centres of waveguides in the lattice
in $x$- and $y$-directions. For meaningful results, please use values around the $10-25~\mu m$ range.   \\ \\
\textsf{Plot vs Plot quickly}    &    These 2 options activate the same procedure for calculating the probability distribution, but display them in different resolutions. \textsf{Plot} gives a figure with approximately 5 times the resolution as \textsf{Plot quickly}.   \\ \\
\textsf{An example}    &    This will show an example system, with $21 \times 21$ nodes all spaced $15~\mu m$ in both $x$ and $y$ directions, and a photon injected in the center (node 11 in both directions). It will then plot the probability distribution of the photon observed at $z = 5~\mu m$ along the waveguides.   \\ 
\end{longtable}

\subsubsection{Input your own positions}
\begin{figure}[htbp]
    \centering
    \includegraphics[scale = 0.47]{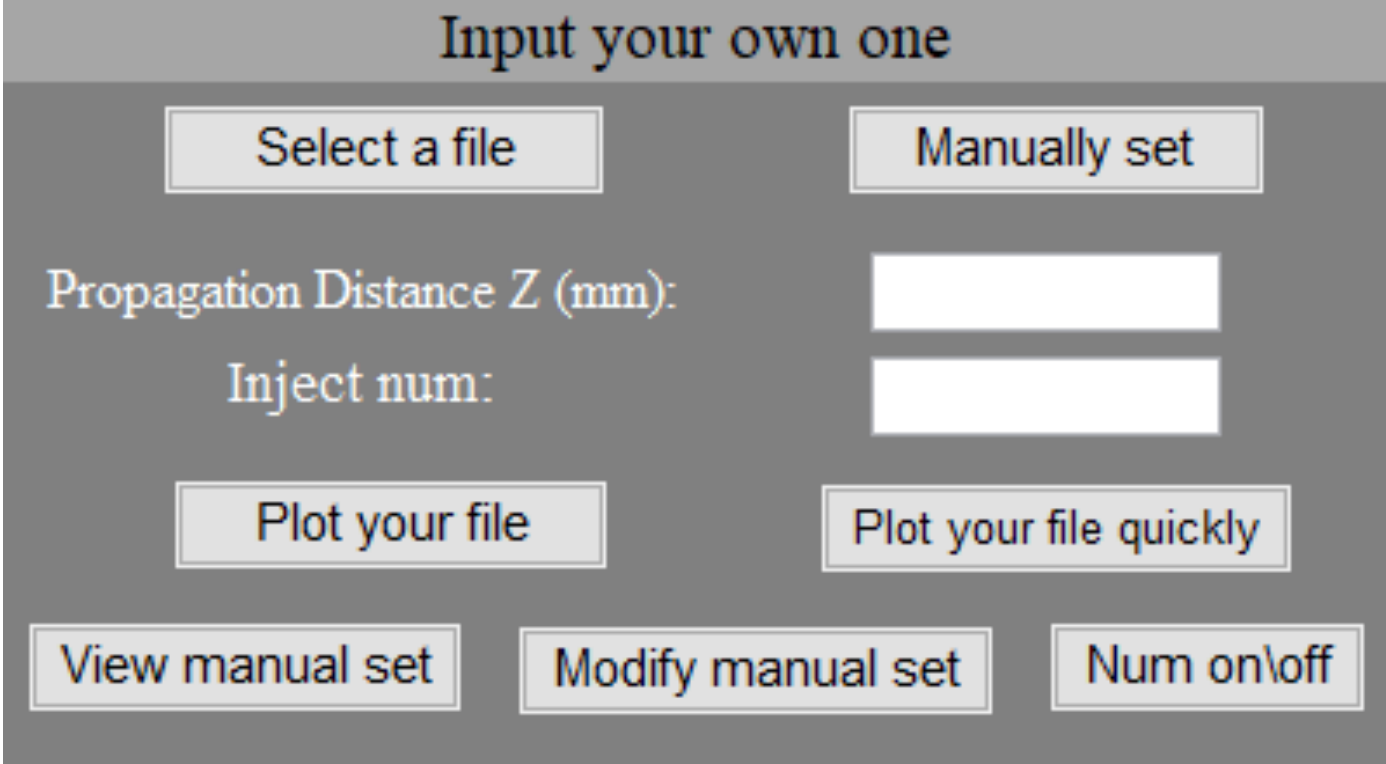}
    \caption{\textbf{The panel for setting a two-dimensional quantum walk with an arbitrary Hamiltonian design.}}
\end{figure}
\begin{longtable}{p{3cm}p{14cm}}
 & This subsection explains the elements on the panel as shown in Fig.~11.\\ \\
\textsf{Select a file}    &    Users can prepare an excel file in the format:\\
\ & \qquad Column 1: label of node (from 1 to n)\\
\ & \qquad Column 2: $x$-coordinate of node\\
\ & \qquad Column 3: $y$-coordinate of node\\
\ & This file can then be imported through this button.   \\ \\
\textsf{Manually set}    &    This clears the workspace and allows the user to set points using their cursor. Left clicking sets a new point at the coordinates of the cursor, while right
clicking removes the last point. To aid in accuracy, the coordinates cursor position corresponds to are displayed in the top left corner.   \\ \\
\textsf{Modify manual set}    &    Allows user to modify an existing set of waveguide positions without clearing the workspace.   \\ \\
\textsf{View manual set}    &    Allow viewing of the exact coordinates that the nodes have been placed at in a list format.   \\ \\
\textsf{Inject no.}    &    The nodes are numbered in the order of listing or manual placement. The point at which a photon enters the array can be set by entering the number of the desired node.   \\ \\
\textsf{Num on/off}    &    This button toggles whether the waveguide numbers are displayed in the workspace. For example, the user might find it useful to toggle \textsf{Num on} when determining the inject point of the photon, or to toggle \textsf{Num off} when it is desirable to reduce clutter.   \\ \\
\    &    The \textsf{Propagation distance}, \textsf{Plot}, and \textsf{Plot quickly} buttons have the same functions as the corresponding ones in the previous section.   \\ \\
\end{longtable}
\subsubsection{Data Output}
\begin{longtable}{p{3cm}p{14cm}}
\    &    There are a few options for exporting data, from the panel on the left part of the GUI. All these types of data can be saved after a file dialogue pops up upon left clicking:   \\ \\
\textsf{Results}    &    This option exports an excel file with the first column containing the probabilities of the photon being in the respective waveguides.   \\ \\
\textsf{Positions}    &    This option outputs an excel file listing positions of the different waveguides. The format is the same as in the file input for waveguide positions (\textsf{Select a file}).   \\ \\
\textsf{Hamiltonian}    &    This option outputs an excel array which lists values of the Hamiltonian matrix.   \\ \\
\textsf{\leftline{Picture output} w/ color bar}    &    This produces a figure which includes the entire panel. Note that since this is a built-in MATLAB function, it will produce higher resolution results than if a print screen command were used.   \\ \\
\textsf{\leftline{Picture output} w/o color bar}    &    This outputs the probability distribution figure by itself.   \\ \\
\end{longtable}
\subsubsection{Other options}
\begin{figure}[htbp]
    \centering
    \includegraphics[scale = 0.8]{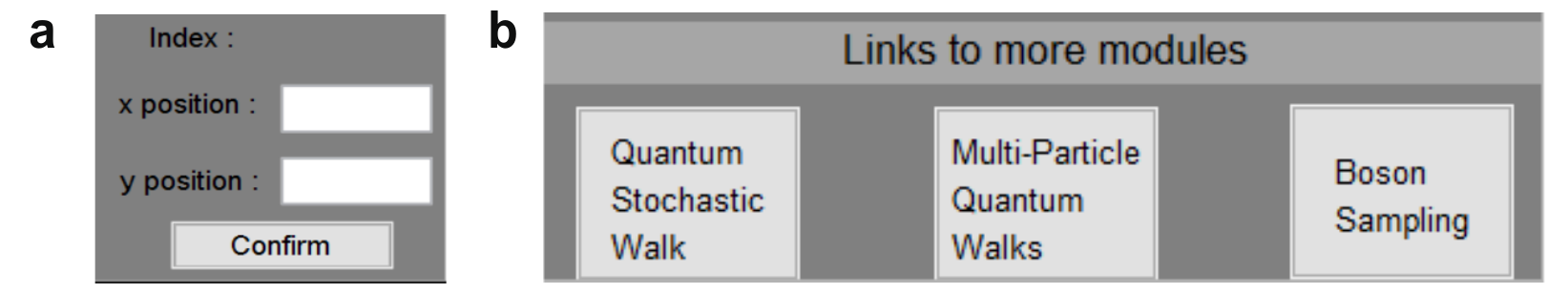}
    \caption{\textbf{Panels for more options.} ({\bf a}) The panel that allows for modifying the position of the selected node shown in the interactive board. ({\bf b}) The panel for links to other modules of FeynmanPAQS.}
\end{figure}
\begin{longtable}{p{3cm}p{14cm}}
\ & More elements in other panels (Fig.~12) are explained as follows:\\ \\
\textsf{Index}    &   the order of the selected waveguides(nodes).    \\  \\
\textsf{x(y) position}    &    show the $x$($y$) position of the selected nodes, which could be modified here.   \\ \\
\textsf{Confirm}    &    apply the $x$($y$) position   \\ \\

\textsf{Colour bar}    &    Our program allows you to choose the colour scheme used to render the probability distribution figure.\\ \\ 
\textsf{\leftline{Links to other} \quad modules}    &    Links to the other modules (\textsf{two-dimensional quantum stochastic walks},
\textsf{multi-particle quantum walks}, and \textsf{boson sampling}) can be found in the
bottom right corner.   \\
\end{longtable}

\subsection{\normalsize Two-dimensional Quantum Stochastic Walks}
\subsubsection{An overview of the user interface}
\begin{figure}[htbp]
    \centering
    \includegraphics[scale = 0.35]{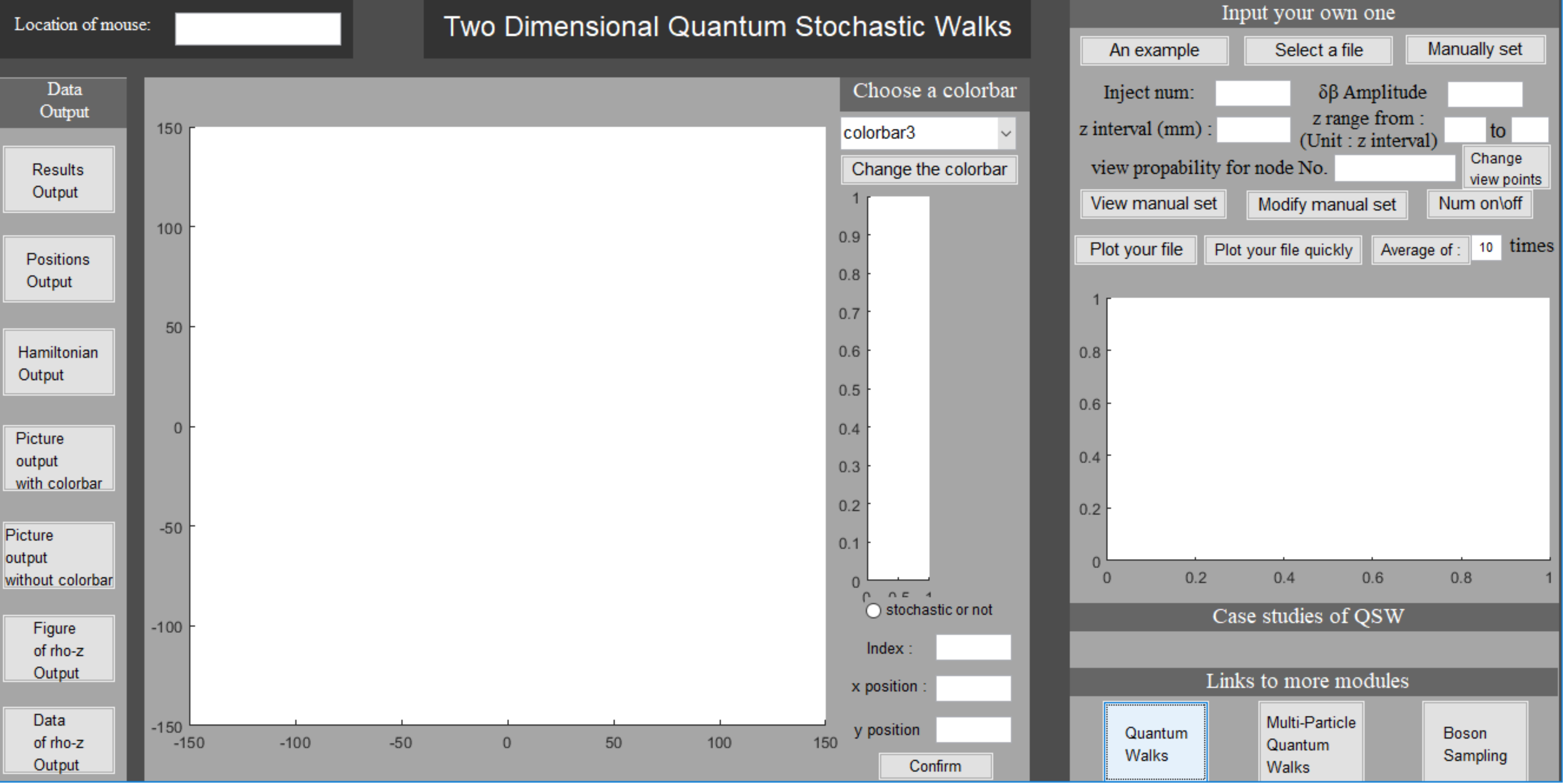}
    \caption{\textbf{The full GUI for the module of QSW before any settings.}}
\end{figure}
The fields in this module (see Fig.~13) are mostly the same as those of the last module. However, there are a few more parameters to set (due to the stochastic nature of the walk), one more plotting option, and one more figure type. This figure is beneath the Manual input panel, and shows the probabilities of selected states (nodes) as a function of $z$, i.e. as the photon propagates. We will call this the $\rho_z$ graph.
\subsubsection{Manual input}
\begin{figure}[htbp]
    \centering
    \includegraphics[scale = 0.5]{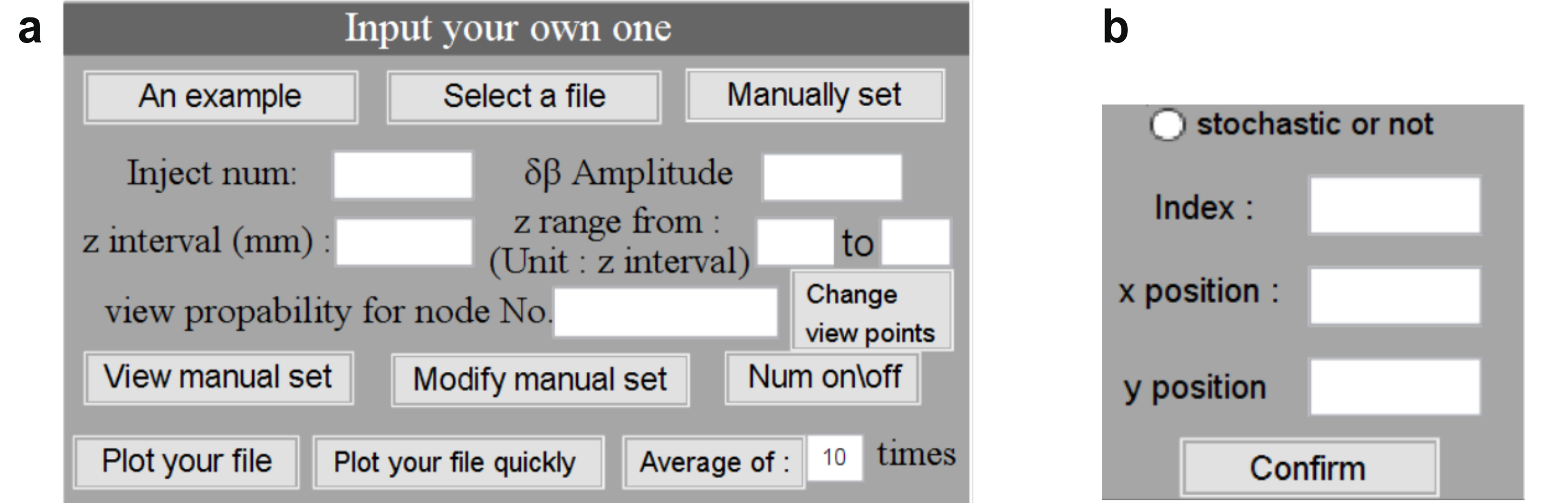}
    \caption{\textbf{{Panels in the module of QSW.}} ({\bf a}) The panel for setting the $\Delta \beta$ model for two-dimensional quantum stochastic walks. ({\bf b}) An additional panel allowing for modifying the position and defining whether or not to apply $\Delta \beta$ for a selected waveguide.} 
\end{figure}
\begin{longtable}{p{3cm}p{14cm}}
\ & This subsection explains the elements on the panel as shown in Fig.~14.\\ \\
\textsf{$\delta \beta$ Amplitude}    &    This is the amplitude of random variation of $\beta$. In practice, this is the number by which we multiply a randomly generated $\delta \beta$ before adding to the previous $\beta$. Experimentally applicable values of $\delta \beta$ range from 0 to about 1.2~mm$^{-1}$.   \\ \\
\textsf{z interval}    &    This is the length by which we keep evolving with the same $\beta$, before we invoke a random change $\delta \beta$ in the Hamiltonian.   \\ \\
\textsf{z range}    &    This is the range of $z$ over which the $\rho_z$ graph is plotted. Note that the initial value does not determine when we start varying $\beta$ (we do this at regular intervals starting from $z$ = 0). However, the final value of $z$ is the value at which we stop the evolution and render the $\rho_z$ graph.   \\ \\
\textsf{\leftline{View probability for} node No.}    &    Here, the user can input which node probabilities are to be plotted in the $\rho_z$ graph. Multiple nodes can be entered. The nodes are, as in the previous module, numbered by order of listing in the input file, or by order of manual input.   \\ \\
\textsf{Average of n times}    &    This option will make the program simulate the quantum stochastic walks for n times and average the results. This applies to both the final probability distribution, and the $\rho_z$ graph.   \\ \\
\textsf{An example}    &    This will show an example system, with a square array of $5 \times 5$ nodes all spaced $12~\mu m$ in both $x$ and $y$ directions, and a photon injected in the centre (node 13). It will then evolve, with $\beta$ changing by an amplitude of $\delta\beta_{max} = 1$~mm$^{-1}$ every 0.1 mm. It will then plot the probability distribution of the photon observed at $z = 5~\mu m$ along the waveguides, and also track the
probability of the photon being in the centre and the top left corner (node 1) from $z$ = 2~mm to 5~mm.   \\ \\
\textsf{Manually set} &  This function is the same as the module of QW, however, here, we use blue circle to represent a node whose $\beta$ will not change and red circle to represent a node whose $\beta$ will change(stochastic). When a node is selected by \textsf{shift + left click}, you can reset by clicking \textsf{stochastic or not}, then \textsf{Confirm} in the panel below the colorbar (Fig.~14.b).  \\ \\
\end{longtable}

\subsection{\normalsize Multi-particle Quantum Walks}
\subsubsection{An overview of the user interface}
\begin{figure}[htbp]
    \centering
    \includegraphics[scale = 0.35]{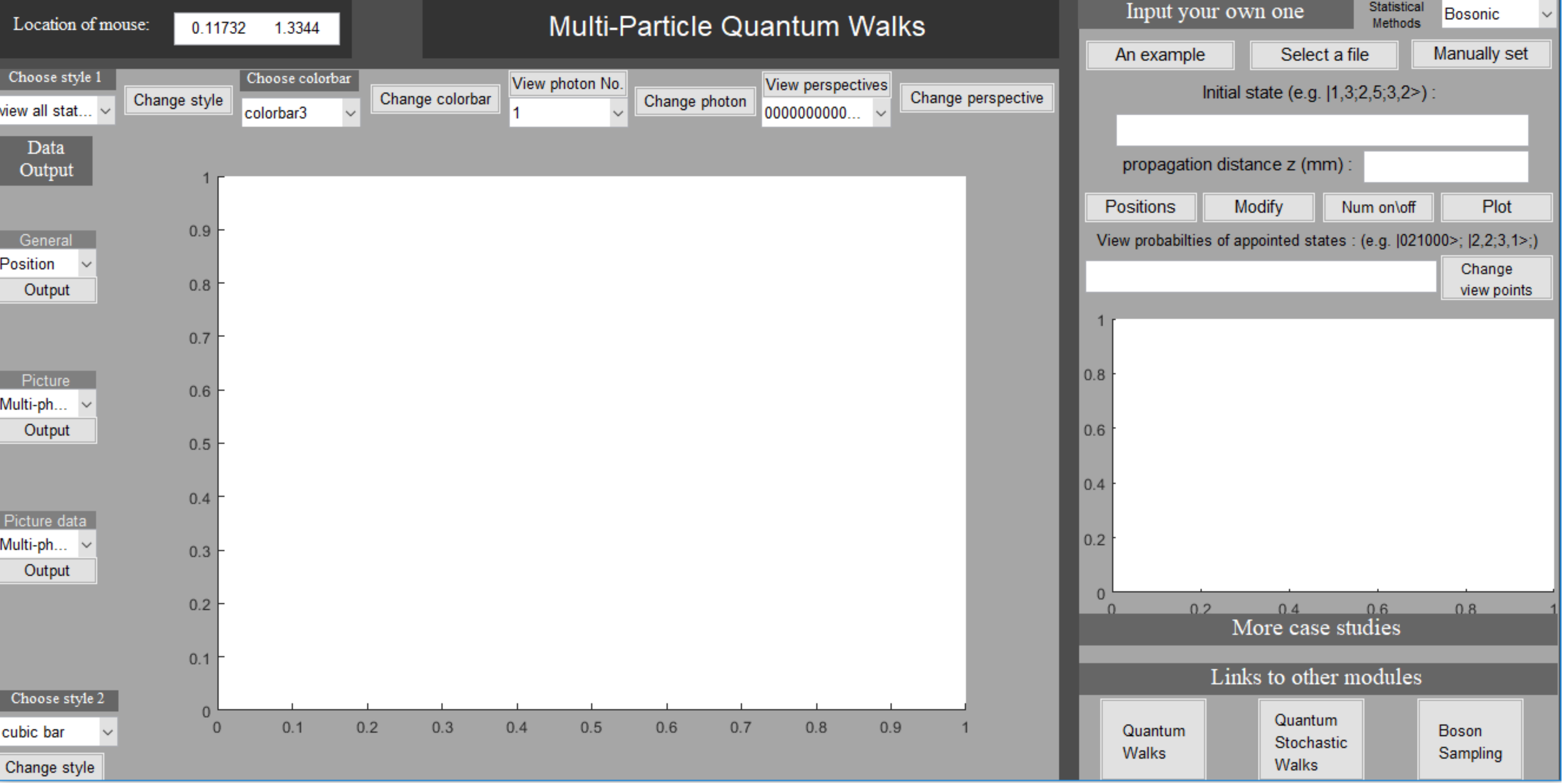}
    \caption{\textbf{The full GUI for the module of MultiParticle before any settings.}}
\end{figure}
This GUI (see Fig.~15) is rather different from the former two. Here, we focus on the probability of quantum correlation, provide different particle types including distinguishable particles and indistinguishable particles with bosonic or fermionic statistics, and change the input style of inject states and view points. Besides, main functions provided in the former two panels are reserved, including \textsf{Manually set}, \textsf{Select a file}, \textsf{Positions}(view manual set), \textsf{Modify}(modify manual set), \textsf{Num on\textbackslash off}, \textsf{Plot} and \textsf{Change view points}.
\subsubsection{Input your own one}
\begin{figure}[htbp]
    \centering
    \includegraphics[scale = 0.3]{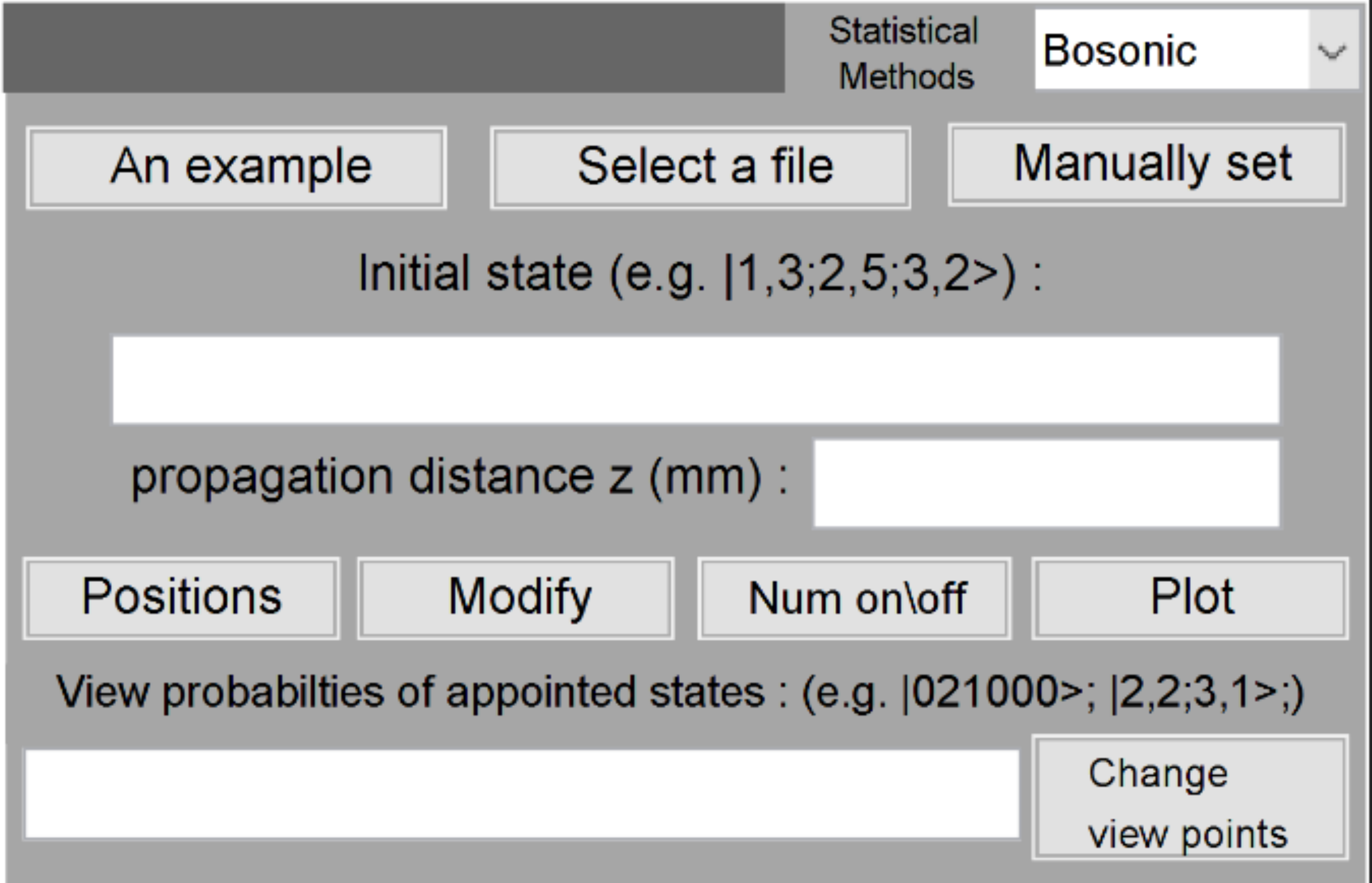}
    \caption{\textbf{The panel for setting parameters for multi-particle quantum walks.}}
\end{figure}

\begin{longtable}{p{3cm}p{14cm}}
\ & This subsection explains the elements on the panel as shown in Fig.~16.\\ \\
\textsf{Statistical Methods}    &    The user can select how the particles evolve in the upper right corner: \textsf{Distinguishable}, or indistinguishable with \textsf{Bosonic} or \textsf{Fermionic} statistics. The default is set to \textsf{Bosonic}.   \\ \\
\textsf{An example}    &    This will show an example system, with 9 nodes in a line spaced 10$\sqrt{2}~\mu m$ and three photons injected in the state of $|100010001>$. It will evolve with bosonic statistic until $z$ = 10~mm along the waveguides. The distribution of probabilities of all states will be plotted first, then the two-photon coorelation in cubic bars (assuming that one photon came out from node No. 1), the picture of facula (assuming that only one photon inject in node No.1), and the probabilities evolution of states $|000020001>, |3,1;5,1;8,1>$ and $|1,1;8,1;9,1>$ will all be plotted.  \\ \\
\textsf{Initial state}    &   Require the following formats interchangeably:
$|i,S_i;j,S_j;\cdots>$,
where $S_i$ is the number of photons in waveguide $i$, or
$|S_1\ S_2\ \cdots S_M>$.
Note that we cannot tell which photon in the output state was originally injected into which waveguide, as we assume they are indistinguishable. (In the distinguishable case, there is no need to view these correlated states, as all information about the output state can be obtained by tracking each individual photon.)    \\ \\
\textsf{View probability of appointed states} & The formats are the same as \textsf{Initial states}, except that there could be more than one viewed state.   \\ \\
\end{longtable}
\subsubsection{The picture panels}
\begin{figure}[htbp]
    \centering
    \includegraphics[scale = 0.35]{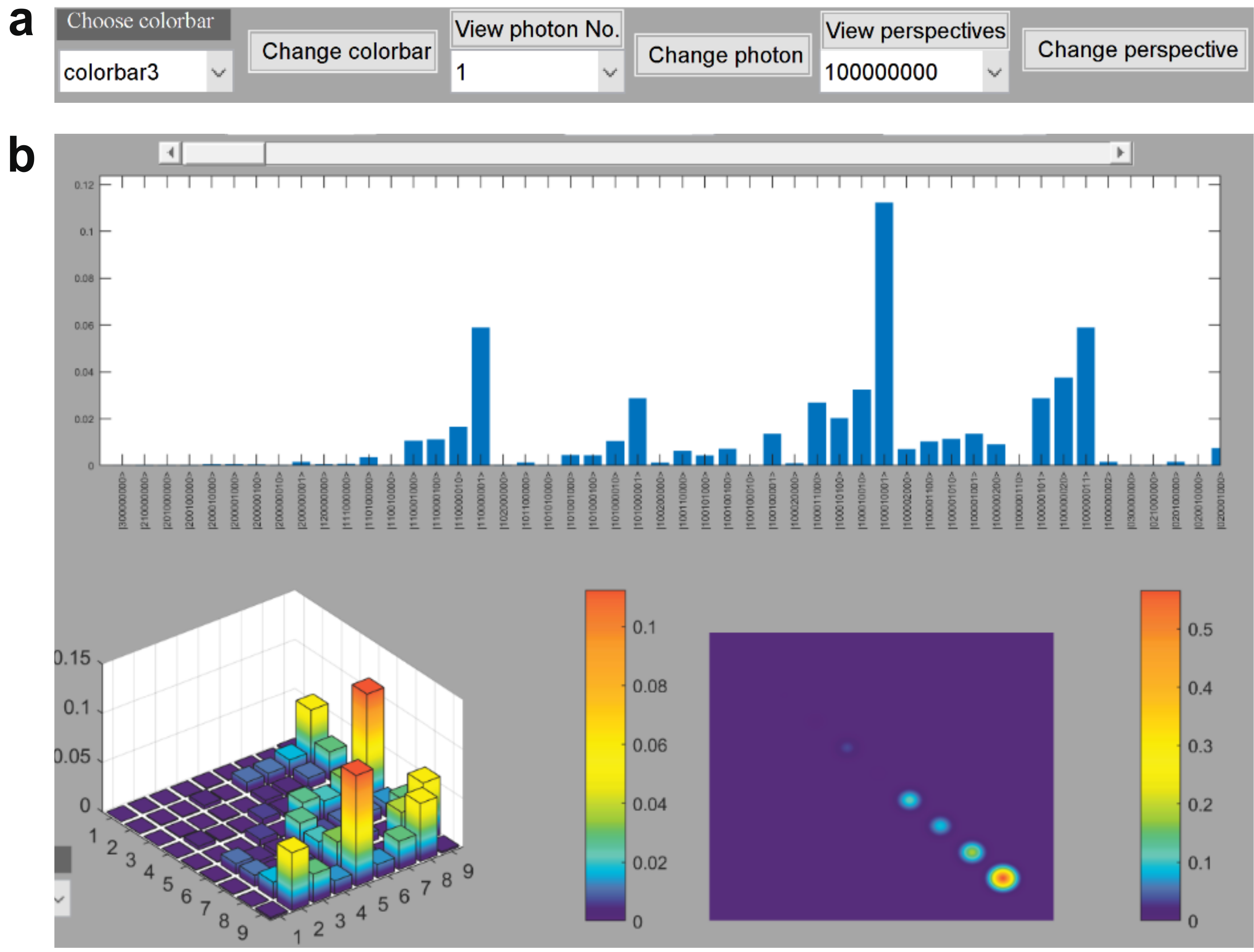}
    \caption{\textbf{{The picture panels.}} ({\bf a}) The panel for setting different perspectives of viewing the figures. ({\bf b}) A screenprint for three picture panels that refer to Multi-particle distribution (the up panel), Two-particle correlation (the bottom-left panel) and Facula of one photon (the bottom-right panel), respectively.}
\end{figure}
\begin{longtable}{p{3cm}p{14cm}}
\ & This subsection explains the elements on the panels as shown in Fig.~17.\\ \\
\textsf{Multi-particle\qquad\quad distribution}    &    A bar graph that shows probabilities for all individual states in descending lexicographical order, i.e. from $|N0\cdots0>$ to $|0\cdots0N>$. If there are more that 16 nodes, then the display will switch from $|30000000>$ to $|1,s_i;etc.>$.  Also note that it will display at most 100 labels for nodes. For labels more than that, it will display with a scroll bar.   \\ \\
\textsf{Two-particle\qquad\quad correlation}    &   This shows the correlations between 2 photons upon fixing the positions of the other N-2 photons. To select the positions of the other photons, select a perspective from \textsf{View perspective} and then click \textsf{Change perspective} that appears on the top of the GUI (Fig.~17a).    \\ \\
\textsf{Facula of one photon}    &    Shows the spatial probability distribution of a single photon after passing through the system, obtained by taking a sum over all correlated states. To change between viewing different photons, select the desired photon under \textsf{View photon No.} and click \textsf{Change photon} (Fig.~17a). THIS ONLY ACCOUNTS FOR SINGLE PHOTONS, DOESN'T DO MULTIPHOTON INTERFERENCE ETC.   \\ \\
\textsf{Probability-propagation\qquad\quad distance graph}    &   A picture panel on the right part of the GUI that allows the user to select specific correlated states and see how their respective probabilities change as the photons propagate along the waveguides. To select which states to plot, enter desired states under \textsf{View probabilities of appointed state} (Fig.~16) in either output state format, and \textsf{Plot}. Note: no matter what range of $z$ is set, there are 100 points for the propagation distance in total.    \\ \\
\end{longtable}

\subsubsection{Detail settings}
\begin{longtable}{p{3cm}p{14cm}}
\textsf{Choose style 1}   &    This pop-up menu that appears on top of the GUI controls the style of \textsf{Multi-photon distribution}. The probability distribution graph displays all states in \textsf{view all states}, whereas a scrollbar appears in \textsf{scroll bar} for closer examination of individual states.    \\ \\
\textsf{Choose style 2}   &    This pop-up menu that appears close to the picture panel of \textsf{Two-photon correlation} controls the style of this 2-photon correlation graph. The graph appears as a 2D array of pixels with colors representing probability amplitudes in \textsf{planar bar}, and as a 3D bar graph in \textsf{cubic bar}.   \\ \\
\end{longtable}
\subsubsection{Data output}
\begin{longtable}{p{3cm}p{14cm}}
\textsf{General}   &   Under the \textsf{General} tab, the position of the waveguides, the Hamiltonian of the system, and probability amplitudes for each photon can be exported in excel format.    \\ \\
\textsf{Picture}   &    The \textsf{Picture} tab can output pictures of the four figures mentioned above, namely, the multi-particle distribution, two-particle correlation, facula of one photon and the probability-propagation distance graph, as well as a view of the full panel, in .png format.    \\ \\
\textsf{Picture Data}   &   The \textsf{Picture data} tab shows options for exporting the arrays used to plot the four types of figures, in excel format.    \\ \\
\end{longtable}

\subsection{\normalsize Boson Sampling}
\subsubsection{An overview of the user interface}
\begin{figure}[htbp]
    \centering
    \includegraphics[scale = 0.35]{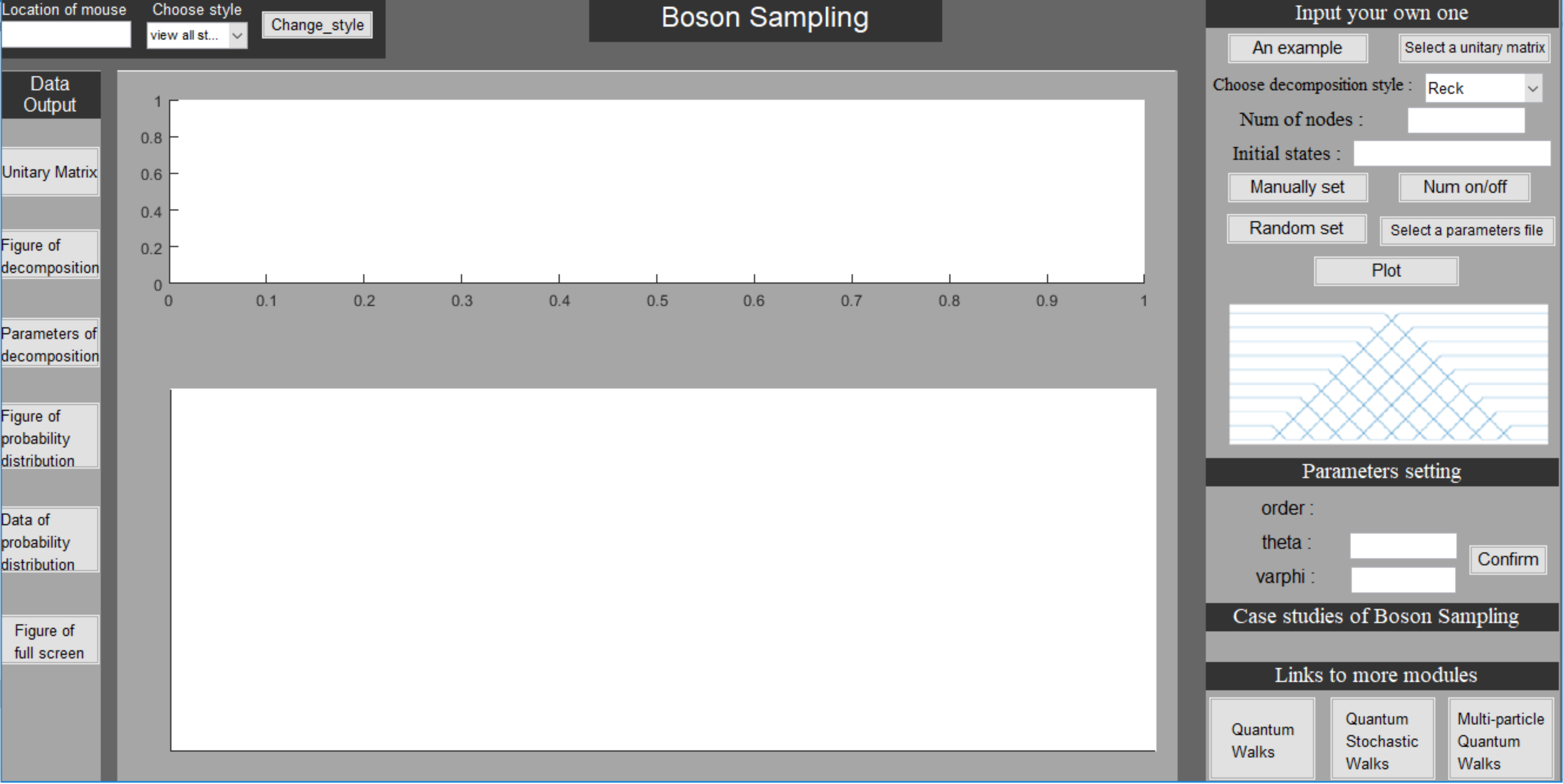}
    \caption{\textbf{The full GUI for the module of BosonSampling before any settings.}}
\end{figure}
This module is similar with the module of MultiParticle. The difference is that in Boson sampling, the positions of waveguides are not variables any more, instead, you need to set the $\theta$ and $\phi$ for each interferometer. Also, propagation distance $z$ do not exit, so there is no picture of $\rho_z$. Besides, several functions are still unchanged, including \textsf{Initial state}, \textsf{Num on\textbackslash off}, \textsf{Choose style}, the \textsf{Parameters setting} panel, and the \textsf{Data output} panel.
\subsubsection{Input your own one}
\begin{figure}[htbp]
    \centering
    \includegraphics[scale = 0.28]{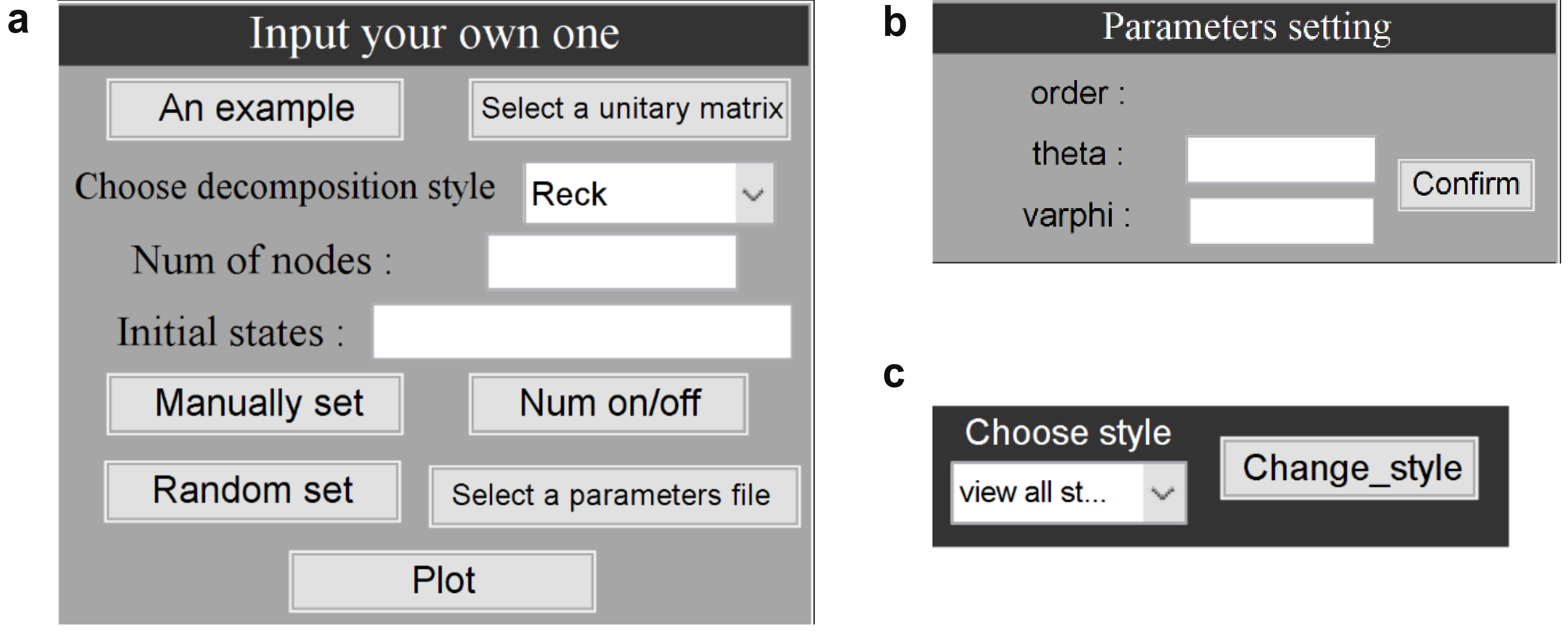}
    \caption{\textbf{{Some panels for the module of BosonSampling.}} ({\bf a}) The panel for setting the unitary scattering matrix, decomposition style and the full decomposition parameter files for boson sampling. ({\bf b}) The panel for setting parameters for each single beam splitter. ({\bf c}) A panel on the top of the GUI allowing for choosing the plotting style for probability distribution. }    
\end{figure}
\begin{longtable}{p{3cm}p{14cm}}
\ & This subsection explains the elements on the panels as shown in Fig.~19.\\ \\
\textsf{An example} &    This will set the number of nodes and the initial state to a default value (number of nodes: 12, initial state : $|000000000111>$), and the decomposition style will set to the default '\textsf{Reck}', while the parameters of the setup will be randomized.    \\ \\
\textsf{Decomposition style} &   As previously described, there are 2 decompositions we use for an arbitrary unitary operator: \textsf{Reck} and \textsf{Clements}.  After chosen, the example picture of the style will appear below the '\textsf{Input your own one}' panel (see Fig.~19a).   \\ \\
\textsf{No. of nodes} &   This is the number of modes, i.e. the boson channels that enter and exit the machine.    \\ \\
\textsf{Manually set} &   If the user has a specific decomposition in mind, then this option clears all parameters and sets all $\theta,\phi=0$. Afterwards, to set parameters for an individual beam splitter (BS), \textsf{Shift + left click} on its position. Its order should show up in the panel named \textsf{Parameters setting} (Fig.~19b). From there, $\theta$ and $\phi$ can be entered. Note that the angles are in units of radians, and input is accepted in both numerical form and in terms of pi (for example, $\frac{\pi}{2}$ would be a valid input, and would show up as 1.57). After inputting values, select confirm and a circle will appear on top of the position of the BS to indicate the input. \\ \\ 
\textsf{Random set} &    This gives each BS a random $\theta$, $\phi$.    \\ \\
\textsf{Num on/off} &    Toggles between whether all parameters are shown on screen or not. Note that it is possible to select whether the parameters for a single BS are shown or not by \textsf{right clicking} on its position.   \\ \\
\textsf{Select a parameter file} &    An excel file can be used as input for the parameters, with format:\\
\ & \qquad column A : order of BS\\
\ & \qquad column B : $\theta$ of BS\\
\ & \qquad column C : $\phi$ of BS  \\ \\ 
\textsf{Select a unitary matrix} &    If a desired unitary transformation is known, an excel file containing complex elements of the matrix can be imported. The program will check for unitarity by checking if $UU^{\dagger}-I=0$, but will not perform a decomposition.   \\ \\
\textsf{Initial state} &  The initial state of photons in channels follows the same format as in previous sections. Additionally, the state is displayed on the left side of the decomposition figure in the interactive board (see Fig.~20), with red markers that can be interacted with to change the number of bosons injected into each channel: \textsf{left clicks} increases, and \textsf{right clicks} decreases. \\ \\
\textsf{Plot} &  After setting up the system, we can \textsf{plot} the \textsf{probability distribution} graph for different correlated states. It can also be displayed in 2 different \textsf{styles}: \textsf{Show all states} or \textsf{scroll bar} (Fig.~19c). \\ \\
\end{longtable}
\begin{figure}[htbp]
    \centering
    \includegraphics[scale = 0.36]{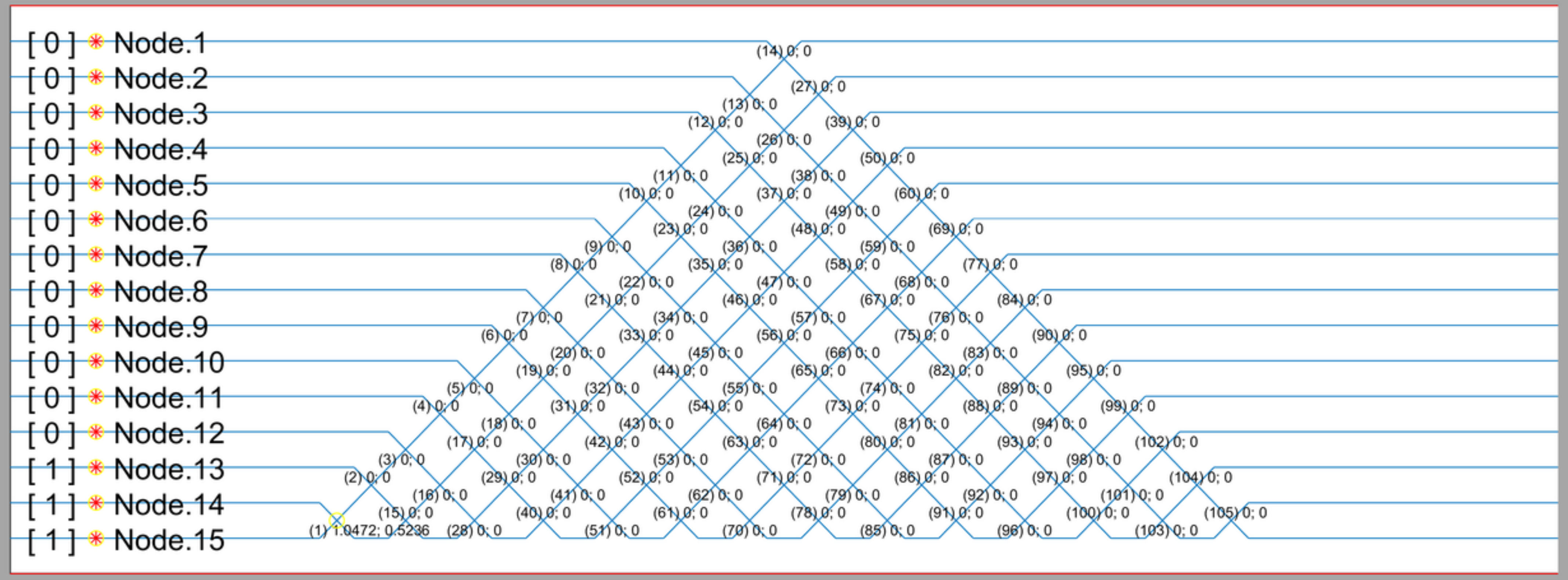}
    \caption{\textbf{The interactive board for setting the BS parameters for boson sampling.}}
\end{figure}
\subsubsection{Data Output}
\begin{longtable}{p{3cm}p{14cm}}
\textsf{Unitary matrix} &    Outputs the unitary transformation matrix with $N\times N$ complex numbers as an excel file.\\ \\
\textsf{Parameters of decomposition} &   Outputs the parameters of the BSs into an excel file, the output format is the same as \textsf{Select a parameter file}.    \\ \\
\end{longtable}


\begin{thebibliography}{10}
\renewcommand{\bibnumfmt}[1]{#1.}

\bibitem{Feynman1982}
\bibinfo{author}{Feynman, R. P.}
\newblock \bibinfo{title}{Simulating Physics with Computers.}
\newblock \emph{\bibinfo{journal}{Int. J. Theor. Phys.}}
 \textbf{\bibinfo{volume}{21}}, \bibinfo{pages}{467-488} (\bibinfo{year}{1982}).

\bibitem{Buluta2009}
\bibinfo{author}{Buluta, I., \& Nori, F.}
\newblock \bibinfo{title}{Quantum simulators.}
\newblock \emph{\bibinfo{journal}{Science}}
  \textbf{\bibinfo{volume}{326}}, \bibinfo{pages}{108-111} (\bibinfo{year}{2009}).

\bibitem{Georgescu2014}
\bibinfo{author}{Georgescu, I. M., Ashhab, S., \& Nori, F.}
\newblock \bibinfo{title}{Quantum simulation.}
\newblock \emph{\bibinfo{journal}{Rev. Mod. Phys.}}
  \textbf{\bibinfo{volume}{86}}, \bibinfo{pages}{153-185} (\bibinfo{year}{2014}).

\bibitem{Arguelloluengo2018}
\bibinfo{author}{Arg\"uelloluengo, J., Gonz\'aleztudela, A., Shi, T., Zoller, P., \& Cirac, J. I.}
\newblock \bibinfo{title}{Analog quantum chemistry simulation.}
\newblock \emph{\bibinfo{journal}{arXiv Preprint}}
  \textbf{\bibinfo{volume}{}}, \bibinfo{pages}{arXiv:1807.09228} (\bibinfo{year}{2018}).

\bibitem{Lambert2013}
\bibinfo{author}{Lambert, N., Chen, Y. N., Cheng, Y. C., Li, C. M., Chen, G. Y., \& Nori, F. }
\newblock \bibinfo{title}{Quantum biology.}
\newblock \emph{\bibinfo{journal}{Nat. Phys.}}
\textbf{\bibinfo{volume}{9}}, \bibinfo{pages}{10-18}
(\bibinfo{year}{2013}).

\bibitem{Flamini2018}
\bibinfo{author}{Flamini, F., Spagnolo, N., \& Sciarrino, F.}
\newblock \bibinfo{title}{Photonic quantum information processing: a review.}
\newblock \emph{\bibinfo{journal}{arXiv Preprint}}
  \textbf{\bibinfo{volume}{}}, \bibinfo{pages}{arXiv:1803.02790v2} (\bibinfo{year}{2018}).

\bibitem{OBrien2010}
\bibinfo{author}{O'Brien, J. L., Furusawa, A., \& Vu\v{c}kovi\'c, J.}
\newblock \bibinfo{title}{Photonic quantum technologies.}
\newblock \emph{\bibinfo{journal}{Nat. Photon.}}
  \textbf{\bibinfo{volume}{3}}, \bibinfo{pages}{687-695} (\bibinfo{year}{2010}).

\bibitem{Aspuru2012}
\bibinfo{author}{Aspuru-Guzik, A., \& Walther, P.}
\newblock \bibinfo{title}{Photonic quantum simulators.}
\newblock \emph{\bibinfo{journal}{Nat. Phys.}}
  \textbf{\bibinfo{volume}{8}}, \bibinfo{pages}{285-291} (\bibinfo{year}{2012}).

\bibitem{Tang2018}
\bibinfo{author}{Tang, H., Lin, X. F., Feng, Z., Chen, J. Y., Gao, J., Sun, K., Wang, C. Y., Lai, P. C., Xu, X. Y., Wang, Y., Qiao, L. F., Yang, A. L., \& Jin, X. M.}
\newblock \bibinfo{title}{Experimental two-dimensional quantum walk on a photonic chip.}
\newblock \emph{\bibinfo{journal}{Sci. Adv.}}
\textbf{\bibinfo{volume}{4}}, \bibinfo{pages}{eaat3174}
(\bibinfo{year}{2018}).

\bibitem{Tang2018b}
\bibinfo{author}{Tang, H., Di Franco, C., Shi, Z. Y., He, T. S., Feng, Z., Gao, J., Li, Z. M., Jiao Z. Q., Wang, T. Y., Kim, M. S.,\& Jin, X. M.}
\newblock \bibinfo{title}{Experimental quantum fast hitting on hexagonal graphs.}
\newblock \emph{\bibinfo{journal}{arXiv Preprint}}
\textbf{\bibinfo{volume}{}}, \bibinfo{pages}{arXiv:1807.06625}
(\bibinfo{year}{2018}).

\bibitem{Harrow2017}
\bibinfo{author}{Harrow, A. W., \& Montanaro, A.}
\newblock \bibinfo{title}{Quantum computational supremacy.}
\newblock \emph{\bibinfo{journal}{Nature}}
\textbf{\bibinfo{volume}{549}}, \bibinfo{pages}{203-209}
(\bibinfo{year}{2017}).

\bibitem{Spring2013}
\bibinfo{author}{Spring, J. B., Metcalf, B. J., Humphreys, P. C., Kolthammer, W. S., Jin, X. M., Barbieri, M., Datta, A., Thomas-Peter, N., Langford, N. K., Kundys, D., Gates, J. C., Smith, B. J., Smith, P. G. R., \& Walmsley, I. A.}
\newblock \bibinfo{title}{Boson sampling on a photonic chip.}
\newblock \emph{\bibinfo{journal}{Science}}
\textbf{\bibinfo{volume}{339}}, \bibinfo{pages}{798-801}
(\bibinfo{year}{2013}).

\bibitem{Broome2013}
\bibinfo{author}{Broome, M.A., Fedrizzi, A., Rahimi-Keshari, S., Dove, J., Aaronson, S., Ralph, T. C., \& White, A. G.}
\newblock \bibinfo{title}{Photonic boson sampling in a tunable circuit.}
\newblock \emph{\bibinfo{journal}{Science}}
\textbf{\bibinfo{volume}{339}}, \bibinfo{pages}{794-798}
(\bibinfo{year}{2013}).

\bibitem{Tillmann2013}
\bibinfo{author}{Tillmann, M., Daki\'c, B., Heilmann, R., Nolte, S., Szameit, A., \& Walther, P.}
\newblock \bibinfo{title}{Experimental boson sampling.}
\newblock \emph{\bibinfo{journal}{Nat. Photon.}}
\textbf{\bibinfo{volume}{7}}, \bibinfo{pages}{540-544}
(\bibinfo{year}{2013}).

\bibitem{Spagnolo2014}
\bibinfo{author}{Spagnolo, N., Vitelli, C., Bentivegna, M., Brod, D. J., Crespi, A., Flamini, F., Giacomini, S., Milani, G., Ramponi, R., Mataloni, P., Osellame, R., Galv\~ao, E. F., \& Sciarrino, F.}
\newblock \bibinfo{title}{Experimental validation of photonic boson sampling.}
\newblock \emph{\bibinfo{journal}{Nat. Photon.}}
\textbf{\bibinfo{volume}{8}}, \bibinfo{pages}{615-620}
(\bibinfo{year}{2014}).

\bibitem{Wang2017}
\bibinfo{author}{Wang, H., He, Y., Li, Y. H., Su, Z. E., Li, B., Huang, H. L., Ding, X., Chen, M. C., Liu, C., Qin, J., Li, J. P., He, Y. M., Schneider, C., Kamp, M., Peng, C. Z., H\"ofling, S., Lu, C. Y. \& Pan, J. W.}
\newblock \bibinfo{title}{High-efficiency multiphoton boson sampling.}
\newblock \emph{\bibinfo{journal}{Nat. Photon.}}
\textbf{\bibinfo{volume}{11}}, \bibinfo{pages}{361-365}
(\bibinfo{year}{2017}).


\bibitem{Ying2018}
\bibinfo{title} {A point in Ying, M. S.'s talk in the seminar `Quantum software: from theory to reality' held in Institute of Software of Chinese Academy of Sciences during Aug. 27-29, 2018.}

\bibitem{Finke2018}
\bibinfo{author} {Finke, D.}
\bibinfo{title} {Quantum computing report. Retrieved from: https://quantumcomputingreport.com/resources/education/.}
(\bibinfo{year}{2018}).

\bibitem{Quantiki2018}
\bibinfo{author}{}
\bibinfo{title}{Quantiki: List of QC simulators. Retrieved from: https://www.quantiki.org/wiki/list-qc-simulators.}
(\bibinfo{year}{2018}).

\bibitem{LaRose2018}
\bibinfo{author} {LaRose, R.}
\bibinfo{title}{Overview and comparison of gate level quantum software platforms.}
\newblock \emph{\bibinfo{journal}{arXiv preprint}}
\textbf{\bibinfo{volume}{}}, \bibinfo{pages}{arXiv:1807.02500}
(\bibinfo{year}{2018}).

\bibitem{Svore2006}
\bibinfo{author} {Svore, K. M., Aho, A. V., Cross, A. W., Chuang, I., \& Markov, I. L.}
\bibinfo{title}{A layered software architecture for quantum computing design tools.}
\newblock \emph{\bibinfo{journal}{Computer}}
\textbf{\bibinfo{volume}{39}}, \bibinfo{pages}{74-83}
(\bibinfo{year}{2006}).

\bibitem{Wecker2014}
\bibinfo{author} {Wecker, D., \& Svore, K. M.}
\bibinfo{title}{L$|$QUi$|$$>$: A software design architecture and domain-specific language for quantum computing.}
\newblock \emph{\bibinfo{journal}{arXiv preprint}}
\textbf{\bibinfo{volume}{}}, \bibinfo{pages}{arXiv:1402.4467}
(\bibinfo{year}{2014}).

\bibitem{Microsoft2017}
\bibinfo{author} {Microsoft.}
\bibinfo{title} {Microsoft quantum development kit. Retrieved from: https://microsoft.com/quantum.}
(\bibinfo{year}{2017}).

\bibitem{Svore2018}
\bibinfo{author} {Svore, K., Geller, A., Troyer, M., Azariah, J., Granade, C., Heim, B., Kliuchnikov, V., Mykhailova, M., Paz, A., \& Roetteler, M.}
\bibinfo{title}{Q\#: Enabling scalable quantum computing and development with a high-level DSL.}
\newblock \emph{\bibinfo{journal}{arXiv preprint}}
\textbf{\bibinfo{volume}{}}, \bibinfo{pages}{arXiv:1803.00652}
(\bibinfo{year}{2018}).

\bibitem{IBM2018}
\bibinfo{author} {IBM.}
\bibinfo{title} {Quantum information science kit. Retrieved from: https://qiskit.org/.}
(\bibinfo{year}{2018}).

\bibitem{Rigetti2018}
\bibinfo{author}{Rigetti, C., Songhurst, C., Pande, V., Pace, P., \& Fitzgerald, A.}
\bibinfo{title} {Forest: An API for quantum computing in the cloud. Retrieved from: https://www.rigetti.com/index.php/forest.}
(\bibinfo{year}{2018}).

\bibitem{Polloreno2018}
\bibinfo{author}{Polloreno, A., \& Zeng, W.}
\bibinfo{title} {Pyquil license. Retrieved from: github.com/rigetticomputing/pyquil/blob/master/LICENSE
\#L204}
(\bibinfo{year}{2018}).

\bibitem{Steiger2016}
\bibinfo{author} {Steiger, D. S., H$\ddot{a}$ner, T., \& Troyer, M.}
\bibinfo{title}{ProjectQ: An open source software framework for quantum computing.}
\newblock \emph{\bibinfo{journal}{arXiv preprint}}
\textbf{\bibinfo{volume}{}}, \bibinfo{pages}{arXiv:1612.08091}
(\bibinfo{year}{2016}).

\bibitem{Liu2017}
\bibinfo{author} {Liu, S. S., Wang, X., Zhou, L., Guan, J., Li, Y. N., He, Y., Duan, R. Y., \&Ying, M. S.}
\bibinfo{title}{$Q|SI\rangle$: a quantum programming environment.}
\newblock \emph{\bibinfo{journal}{arXiv preprint}}
\textbf{\bibinfo{volume}{}}, \bibinfo{pages}{arXiv:1710.09500}
(\bibinfo{year}{2017}).


\bibitem{IBMcloud2016}
\bibinfo{author} {IBM.}
\bibinfo{title} {IBM Q Experience. Retrieved from:
https://quantumexperience.ng.bluemix.net/qx/experience.}
(\bibinfo{year}{2017}).

\bibitem{Alibaba2017}
\bibinfo{author} {Alibaba.}
\bibinfo{title} {Alibaba's quantum cloud platform. Retrieved from:
http://quantumcomputer.ac.cn/index.html.}
(\bibinfo{year}{2017}).

\bibitem{Xin2018}
\bibinfo{author} {Xin, T., Huang, S., Lu, S., Li, K., Luo, Z.,Yin, Z., Li, J., Lu, D., Long, G., \& Zeng, B.}
\bibinfo{title}{Nmrcloudq: A quantum cloud experience on a nuclear magnetic resonance quantum computer.}
\newblock \emph{\bibinfo{journal}{Science Bulletin}}
\textbf{\bibinfo{volume}{63}}, \bibinfo{pages}{17-23}
(\bibinfo{year}{2018}).

\bibitem{Mcclean2017}
\bibinfo{author} {Mcclean, J. R., Kivlichan, I. D., Sung, K. J., Steiger, D. S., Cao, Y., \& Dai, C., et al.}
\bibinfo{title}{OpenFermion: the electronic structure package for quantum computers.}
\newblock \emph{\bibinfo{journal}{arXiv preprint}}
\textbf{\bibinfo{volume}{}}, \bibinfo{pages}{arXiv:1710.07629}
(\bibinfo{year}{2017}).

\bibitem{Johansson2012}
\bibinfo{author} {Johansson, J. R., Nation, P. D., \& Nori, F.}
\bibinfo{title}{QuTiP: An open-source Python framework for the dynamics of open quantum systems.}
\newblock \emph{\bibinfo{journal}{Comp. Phys. Comm.}}
\textbf{\bibinfo{volume}{183}}, \bibinfo{pages}{1760-1772}
(\bibinfo{year}{2012}).

\bibitem{Johansson2013}
\bibinfo{author} {Johansson, J. R., Nation, P. D., \& Nori, F.}
\bibinfo{title}{QuTiP 2: A Python framework for the dynamics of open quantum systems.}
\newblock \emph{\bibinfo{journal}{Comp. Phys. Comm.}}
\textbf{\bibinfo{volume}{184}}, \bibinfo{pages}{1234}
  (\bibinfo{year}{2013}).

\bibitem{Falloon2017}
\bibinfo{author} {Falloon, P., Rodriguez, J. \& Wang, J.}
\bibinfo{title} {Qswalk: a mathematica, package for quantum stochastic walks on arbitrary graphs.}
\newblock \emph{\bibinfo{journal}{Comput. Phys. Commun.}}
\textbf{\bibinfo{volume}{217}}, \bibinfo{pages}{162-170}
  (\bibinfo{year}{2017}).

\bibitem{Izaac2016}
\bibinfo{author} {Izaac, J.}
\bibinfo{title} {Mathematica package of centrality testing. Retrieved from:http://demonstrations.wolfram.com/PTSymmetric
QuantumWalksAndCentralityTestingOnDirectedGraphs/.}
  (\bibinfo{year}{2016}).

\bibitem{Izaac2017}
\bibinfo{author} {Izaac, J., Wang, J. B., Abbott, P. C., \& Ma, X. S.}
\bibinfo{title} {Quantum centrality testing on directed graphs via PT-Symmetric quantum walks.}
\newblock \emph{\bibinfo{journal}{Phys. Rev. A}}
\textbf{\bibinfo{volume}{96}}, \bibinfo{pages}{032305}
  (\bibinfo{year}{2017}).

\bibitem{Killoran2018}
\bibinfo{author} {Killoran, N., Izaac, J., Quesada, N., Bergholm, V., Amy, M., \& Weedbrook, C.}
\bibinfo{title}{Strawberry Fields: A software platform for photonic quantum computing.}
\newblock \emph{\bibinfo{journal}{arXiv preprint}}
\textbf{\bibinfo{volume}{}}, \bibinfo{pages}{arXiv:1804.03159}
(\bibinfo{year}{2017}).

\bibitem{Bristol2018}
\bibinfo{author}{Bristol University}
\bibinfo{title} {Bristol Quantum Cloud for Boson Sampling. Retrieved from: http://cnotmz.appspot.com/\#}
(\bibinfo{year}{2018}).

\bibitem{Wang2014}
\bibinfo{author}{Wang, J., Santamato, A., Jiang, P., Bonneau, D., Engin, E., Silverstone, J. W., Lermer, M., Beetz, J., Kamp, M., Hofling, S., Tanner, M. G., Natarajan, C. M., Hadfield, R. H., Dorenbos, S. N., Zwiller, V., O'Brien, J. L., \& Thompson, M. G.}
\newblock \bibinfo{title}{Gallium arsenide (GaAs) quantum photonic waveguide circuits.}
\newblock \emph{\bibinfo{journal}{Opt. Commun.}}
  \textbf{\bibinfo{volume}{327}}, \bibinfo{pages}{49-55} (\bibinfo{year}{2014}).

\bibitem{Politi2008}
\bibinfo{author}{Politi, A., Cryan, M. J., Rarity, J. G., Yu, S., \& O'Brien, J. L.}
\newblock \bibinfo{title}{Silica-on-silicon waveguide quantum circuits.}
\newblock \emph{\bibinfo{journal}{Science}}
  \textbf{\bibinfo{volume}{320}}, \bibinfo{pages}{646-649} (\bibinfo{year}{2008}).

\bibitem{Feng2016}
\bibinfo{author}{Feng, Z., Wu, B. H., Zhao, Y. X., Gao, J., Qiao, L. F., Yang, A. L., Lin, X. F., \& Jin, X. M.}
\newblock \bibinfo{title}{Invisibility Cloak Printed on a Photonic Chip.}
\newblock \emph{\bibinfo{journal}{Sci. Rep.}}
\textbf{\bibinfo{volume}{6}}, \bibinfo{pages}{28527} (\bibinfo{year}{2016}).

\bibitem{Szameit2007}
\bibinfo{author}{Szameit, A., Dreisow, F., Pertsch, T., Nolte, S.,\& Trnnermann, A. }
\newblock \bibinfo{title}{Control of directional evanescent coupling in fs laser written waveguides.}
\newblock \emph{\bibinfo{journal}{Opt. Express}}
\textbf{\bibinfo{volume}{15}}, \bibinfo{pages}{1579-1587}
(\bibinfo{year}{2007}).

\bibitem{Crespi2013}
\bibinfo{author}{Crespi, A., Osellame, R., Ramponi, R., Brod, D. J., Galv\~ao, E. F., Spagnolo, N., Vitelli, C., Maiorino, E., Mataloni, P., \& Sciarrino, F.}
\newblock \bibinfo{title}{Integrated multimode interferometers with arbitrary designs for photonic boson sampling. }
\newblock \emph{\bibinfo{journal}{Nat. Photon.}}
\textbf{\bibinfo{volume}{7}}, \bibinfo{pages}{545-549}
(\bibinfo{year}{2013})

\bibitem{Chaboyer2015}
\bibinfo{author}{Chaboyer, Z., Meany, T., Helt, L. G., Withford, M. J., \& Steel, M. J.}
\newblock \bibinfo{title}{Tunable quantum interference in a 3D integrated circuit.}
\newblock \emph{\bibinfo{journal}{Sci. Rep.}}
\textbf{\bibinfo{volume}{5}}, \bibinfo{pages}{9601}
(\bibinfo{year}{2015})

\bibitem{Giorgino2017}
\bibinfo{author}{Giorgino, T., Laio, A., \& Rodriguez, A.}
\newblock \bibinfo{title}{Metagui 3: a graphical user interface for choosing the collective variables in molecular dynamics simulations.}
\newblock \emph{\bibinfo{journal}{Compt. Phys. Commun.}}
\textbf{\bibinfo{volume}{217}}, \bibinfo{pages}{204-209}
(\bibinfo{year}{2017}).

\bibitem{Umansky2012}
\bibinfo{author}{Umansky, M., \& Weihs, D.}
\newblock \bibinfo{title}{Novel algorithm and matlab-based program for automated power law analysis of single particle, time-dependent mean-square displacement.}
\newblock \emph{\bibinfo{journal}{Compt. Phys. Commun.}}
\textbf{\bibinfo{volume}{183}}, \bibinfo{pages}{1783-1792}
(\bibinfo{year}{2012}).

\bibitem{Vergaraperez2016}
\bibinfo{author}{Vergaraperez, S., \& Marucho, M.}
\newblock \bibinfo{title}{Mpbec, a matlab program for biomolecular electrostatic calculations.}
\newblock \emph{\bibinfo{journal}{Compt. Phys. Commun.}}
\textbf{\bibinfo{volume}{198}}, \bibinfo{pages}{179-194}
(\bibinfo{year}{2016}).

\bibitem{Aharonov1993}
\bibinfo{author}{Aharonov, Y., Davidovich, L., \& Zagury, N.}
\newblock \bibinfo{title}{Quantum random walks}.
\newblock \emph{\bibinfo{journal}{Phys. Rev. A}}
  \textbf{\bibinfo{volume}{48}}, \bibinfo{pages}{1687} (\bibinfo{year}{1993}).

\bibitem{Childs2002}
\bibinfo{author}{Childs, A. M., Farhi, E., \& Gutmann, S. }
\newblock \bibinfo{title}{An example of the difference between quantum and classical random walks. }
\newblock \emph{\bibinfo{journal}{Quantum Inf. Process}}
  \textbf{\bibinfo{volume}{1}}, \bibinfo{pages}{35-43} (\bibinfo{year}{2002}).

\bibitem{Ambainis2003}
\bibinfo{author}{Ambainis, A.}
\newblock \bibinfo{title}{Quantum walks and their algorithmic applications.}
\newblock \emph{\bibinfo{journal}{Int. J. Quantum Inf.}}
  \textbf{\bibinfo{volume}{1}}, \bibinfo{pages}{507-518} (\bibinfo{year}{2003}).

\bibitem{Shenvi2003}
\bibinfo{author}{Shenvi, N., Kempe, J., \& Whaley, K. B. }
\newblock \bibinfo{title}{Quantum random-walk search algorithm. }
\newblock \emph{\bibinfo{journal}{Phys. Rev. A}}
  \textbf{\bibinfo{volume}{67}}, \bibinfo{pages}{052307} (\bibinfo{year}{2003}).

\bibitem{Childs2004}
\bibinfo{author}{Childs, A. M., \& Goldstone, J. }
\newblock \bibinfo{title}{Spatial search by quantum walk.}
\newblock \emph{\bibinfo{journal}{Phys. Rev. A}}
  \textbf{\bibinfo{volume}{70}}, \bibinfo{pages}{022314} (\bibinfo{year}{2004}).

\bibitem{Mulken2011}
\bibinfo{author}{M\"ulken, O., \& Blumen, A.}
\newblock \bibinfo{title}{Continuous-time quantum walks: Models for coherent transport on complex networks. }
\newblock \emph{\bibinfo{journal}{Phys. Rep.}}
  \textbf{\bibinfo{volume}{502}}, \bibinfo{pages}{37-87} (\bibinfo{year}{2011}).

\bibitem{Caruso2016}
\bibinfo{author} {Caruso, F., Crespi, A., Ciriolo, A. G., Sciarrino, F., \& Osellame, R.}
\bibinfo{title} {Fast escape of a quantum walker from an integrated photonic maze.}
\newblock \emph{\bibinfo{journal}{Nat. Commun.}}
\textbf{\bibinfo{volume}{7}}, \bibinfo{pages}{11682}
  (\bibinfo{year}{2016}).

\bibitem{Whitfield2010}
\bibinfo{author}{Whitfield, J. D., Rodr\'iguez-Rosario, C. A., \& Aspuru-Guzik, A.}
\newblock \bibinfo{title}{Quantum stochastic walks: A generalization of classical random walks and quantum walks.}
\newblock \emph{\bibinfo{journal}{Phys. Rev. A}}
  \textbf{\bibinfo{volume}{81}}, \bibinfo{pages}{022323} (\bibinfo{year}{2010}).


\bibitem{Hong1987}
\bibinfo{author}{Hong, C. K., Ou, Z. Y., \& Mandel. L.}
\newblock \bibinfo{title}{Measurement of subpicosecond time intervals between two photons by interference.}
\newblock \emph{\bibinfo{journal}{Phys. Rev. Lett.}}
\textbf{\bibinfo{volume}{59}}, \bibinfo{pages}{2044–2046}
(\bibinfo{year}{1987}).

\bibitem{Peruzzo2010}
\bibinfo{author}{Peruzzo, A., Lobino, M., Matthews, J. C. F., Matsuda, N., Politi, A., Poulios, K., Zhou, X.-Q., Lahini, Y., Ismaili, N., Worhoff, K., Bromberg, Y., Silberberg, Y., Thompson, M. G., \& O'Brien, J. L.}
\newblock \bibinfo{title}{Quantum walks of correlated photons}
\newblock \emph{\bibinfo{journal}{Science}}
\textbf{\bibinfo{volume}{329}}, \bibinfo{pages}{1500-1503}
(\bibinfo{year}{2010}).

\bibitem{Gao2016}
\bibinfo{author} {Gao, J., Qiao, L. F., Lin, X. F., Jiao, Z. Q., Feng, Z., Zhou, Z., Gao, Z. W., Xu, X. Y., Chen, Y., Tang, H. \& Jin, X.M.}
\bibinfo{title} {Non-classical photon correlation in a two-dimensional photonic lattice.}
\newblock \emph{\bibinfo{journal}{Opt. Express}}
\textbf{\bibinfo{volume}{24}}, \bibinfo{pages}{12607-12616}
  (\bibinfo{year}{2016}).

\bibitem{Sansoni2012}
\bibinfo{author}{Sansoni, L., Sciarrino, F., Vallone, G., Mataloni, P., Crespi, A., Ramponi, R. \& Osellame, R.}
\newblock \bibinfo{title}{Two-Particle Bosonic-Fermionic Quantum Walk via Integrated Photonics.}
\newblock \emph{\bibinfo{journal}{Phys. Rev. Lett.}}
\textbf{\bibinfo{volume}{108}}, \bibinfo{pages}{010502}
(\bibinfo{year}{2012}).

\bibitem{Matthews2013}
\bibinfo{author}{Matthews, J. C., Poulios, K., Meinecke, J. D., Politi, A., Peruzzo, A., Ismail, N., W\"orhoff, K., Thompson, M. G., \& O'Brien, J. L.}
\newblock \bibinfo{title}{Observing fermionic statistics with photons in arbitrary processes.}
\newblock \emph{\bibinfo{journal}{Sci. Rep.}}
\textbf{\bibinfo{volume}{3}}, \bibinfo{pages}{1719}
(\bibinfo{year}{2013}).

\bibitem{Aaronson2013}
\bibinfo{author}{Aaronson, S. \& Arkhipov, A.}
\newblock \bibinfo{title}{The computational complexity of linear optics.}
\newblock \emph{\bibinfo{journal}{Theory Comput.}}
\textbf{\bibinfo{volume}{9}}, \bibinfo{pages}{143-252}
(\bibinfo{year}{2013}).

\bibitem{Reck1994}
\bibinfo{author} {Reck, M., Zeilinger, A., Bernstein, H. J., \& Bertani, P.}
\bibinfo{title} {Experimental realization of any discrete unitary operator.}
\newblock \emph{\bibinfo{journal}{Phys. Rev. Lett.}}
\textbf{\bibinfo{volume}{73}}, \bibinfo{pages}{58-61}
(\bibinfo{year}{1994}).

\bibitem{Clements2016}
\bibinfo{author} {Clements, W. R., Humphreys, P. C., Metcalf, B. J., Kolthammer, W. S., \& Walmsley, I. A.}
\bibinfo{title} {Optimal design for universal multiport interferometers.}
\newblock \emph{\bibinfo{journal}{Optica}}
\textbf{\bibinfo{volume}{3}}, \bibinfo{pages}{1460-1465}
(\bibinfo{year}{2016}).



\bibitem{Ryser1963}
\bibinfo{author} {Ryser, H. J.}
\newblock \emph{\bibinfo{title}{Combinatorial Mathematics, Vol. 14 of The Carus Mathematical Monographs.}}
\bibinfo{publishing house}{Mathematical Association of America}
(\bibinfo{year}{1963}).

\bibitem{Glynn2010}
\bibinfo{author} {Glynn, D. G.}
\bibinfo{title} {The permanent of a square matrix.}
\newblock \emph{\bibinfo{journal}{European Journal of Combinatorics}}
\textbf{\bibinfo{volume}{31}}, \bibinfo{pages}{1887-1891}
(\bibinfo{year}{2010}).

\bibitem{Glynn2013}
\bibinfo{author} {Glynn, D. G.}
\bibinfo{title} {Permanent formulae from the Veronesean.}
\newblock \emph{\bibinfo{journal}{Designs, Codes and Cryptography}}
\textbf{\bibinfo{volume}{68}}, \bibinfo{pages}{39-47}
(\bibinfo{year}{2013}).

\bibitem{Nijenhuis1978}
\bibinfo{author} {Nijenhuis, A., \&Wilf, H. S.}
\newblock \emph{\bibinfo{journal}{Combinatorial algorithms: for computers and calculators, 2nd ed.}}
\bibinfo{publication house}{New York: Academic Press}
(\bibinfo{year}{1978}).


\end{thebibliography}
\end{document}